\documentclass[fleqn,usenatbib]{mnras}

\usepackage[T1]{fontenc}

\DeclareRobustCommand{\VAN}[3]{#2}
\let\VANthebibliography\thebibliography
\def\thebibliography{\DeclareRobustCommand{\VAN}[3]{##3}\VANthebibliography}

\usepackage{graphicx}	
\usepackage{amsmath}	
\usepackage{float}
\usepackage{natbib}
\usepackage{caption}
\usepackage{gensymb}
\usepackage{bm}
\usepackage[output-decimal-marker={,},exponent-product=\cdot]{siunitx}
\usepackage[utf8]{inputenc}
\usepackage{hyperref}

\newcommand{\mr}{\mathrm}

\usepackage{newtxtext,newtxmath}

\title[Spin-driven jet feedback]{Spin-driven jet feedback in idealised simulations of galaxy groups and clusters}

\author[F. Huško et al.]{
Filip Huško,$^{1}$\thanks{E-mail: filip.husko@durham.ac.uk}
Cedric G. Lacey,$^{1}$ 
Joop Schaye,$^{2}$ 
Matthieu Schaller,$^{2,3}$ 
Folkert S.J. Nobels,$^{2}$ 
\\
$^{1}$ Institute for Computational Cosmology, Department of Physics, University of Durham, South Road, Durham, DH1 3LE, UK\\
$^{2}$ Leiden Observatory, Leiden University, PO Box 9513, NL-2300 RA Leiden,
the Netherlands\\
$^{3}$ Lorentz Institute for Theoretical Physics, Leiden University, PO box 9506, 2300 RA Leiden, the Netherlands
}

\date{Accepted XXX. Received YYY; in original form ZZZ}

\pubyear{2022}

\begin{document}
\label{firstpage}
\pagerange{1--23}
\maketitle

\begin{abstract}
We implement a black hole spin evolution and jet feedback model into SWIFT, a smoothed particle hydrodynamics code. The jet power is determined self-consistently assuming that the black hole accretion rate is equal to the Bondi rate (i.e. the accretion efficiency is $100\%$), and using a realistic, spin-dependent efficiency. The jets are launched along the spin axis of the black hole, resulting in natural reorientation and precession. We apply the model to idealised simulations of galaxy groups and clusters, finding that jet feedback successfully quenches gas cooling and star formation in all systems. Our group-size halo ($M_\mr{200}=10^{13}$ $\mr{M}_\odot$) is quenched by a strong jet episode triggered by a cooling flow, and it is kept quenched by a low-power jet fed from hot halo accretion. In more massive systems ($M_\mr{200}\gtrsim10^{14}$ $\mr{M}_\odot$), hot halo accretion is insufficient to quench the galaxies, or to keep them quenched after the first cooling episode. These galaxies experience multiple episodes of gas cooling, star formation and jet feedback. In the most massive galaxy cluster that we simulate ($M_\mr{200}=10^{15}$ $\mr{M}_\odot$), we find peak cold gas masses of $10^{10}$ $\mr{M}_\odot$ and peak star formation rates of a few times $100$ $\mr{M}_\odot\mr{yr}^{-1}$. These values are achieved during strong cooling flows, which also trigger the strongest jets with peak powers of $10^{47}$ $\mathrm{erg}\hspace{0.3mm}\mathrm{s}^{-1}$. These jets subsequently shut off the cooling flows and any associated star formation. Jet-inflated bubbles draw out low-entropy gas that subsequently forms dense cooling filaments in their wakes, as seen in observations.
\end{abstract}


\begin{keywords}
galaxies: jets -- galaxies: evolution -- galaxies: intracluster medium
\end{keywords}




\section{Introduction}

Observations of massive elliptical galaxies reveal that they are mostly 'red and dead', i.e. devoid of significant amounts of star-forming gas and young stars. With the exception of a minority of brightest cluster galaxies (hereafter BCGs, \citealt{Edge2001}), these ellipticals host small amounts of cold atomic and molecular gas ($<1\%$ in terms of gas-to-stellar mass fraction; e.g. \citealt{Wiklin1995}, \citealt{Young2011}, \citealt{Davis2019}) are almost completely devoid of warm ionised gas (e.g. \citealt{Phillips1986}, \citealt{Morganti2006}, \citealt{Temi2022}), and therefore host little ongoing star formation (e.g. \citealt{Salim2007}, \citealt{Whitaker2012}). The typical stellar ages in these galaxies imply that most of the stars formed more than several Gyr ago (e.g. \citealt{Bell2004}, \citealt{vanDokkum2010}). Theoretical models of galaxy formation, be they semi-empirical (e.g. \citealt{Behroozi2013}, \citealt{Moster2018}), semi-analytic (e.g. \citealt{Bower2006}, \citealt{Somerville2008}, \citealt{Henriques2015}, \citealt{Lacey2016}), or in the form of hydrodynamical simulations (e.g. \citealt{Dubois2014}, \citealt{Vogelsberger2014}, \citealt{Schaye2015}), find that energy injection from active galactic nuclei (AGN feedback) powered by supermassive black holes (SMBHs) at the centres of massive galaxies is required in order to produce such quenching of star formation (e.g. \citealt{Croton2006}).

X-ray observations of hot gaseous haloes around galaxies reveal evidence of AGN feedback in the form of cavities in the X-ray emitting gas (\citealt{Boehringer1993}, \citealt{McNamara2005}, \citealt{Wise2007}). These observations have focused mostly on massive galaxy clusters (with dark matter halo masses of $M_\mr{200}\simeq10^{15}$ $\mr{M}_\odot$), mostly due to the large X-ray luminosity of the intracluster medium (\citealt{Sarazin1986}, \citealt{Stanek2006}). Group-size gas haloes also display such cavities in the intergalactic medium (\citealt{Birzan2004}, \citealt{Eckert2021}), despite being harder to detect. Observations at radio frequencies often find that these X-ray cavities are coincident with lobes (bubbles) of synchrotron-emitting plasma whose source is the central SMBH of the central galaxy (\citealt{Biermann1987}, \citealt{Odea1998}, \citealt{Markoff2001}). This plasma originates from jets of relativistic particles launched from the vicinities of SMBHS (\citealt{Blandford1979}, \citealt{Urry1995}).

The properties of X-ray cavities can be used to estimate the jet powers required to inflate them (\citealt{Gull1973}, \citealt{Churazov2000}, \citealt{Fabian2012}, \citealt{Werner2019}, \citealt{Eckert2021}). Such analyses indicate that the cavity (jet) powers are correlated with the X-ray luminosities of the gaseous atmospheres, both in galaxy groups and clusters (\citealt{Rafferty2006}, \citealt{Hlavacek-Larrondo2012}, \citealt{Russell2013}). This suggests that the SMBHs are fed from the gas that cools from those atmospheres, since the X-ray luminosity of such gas can be connected to its cooling and inflow rates (\citealt{White1997}, \citealt{Peres1998}). The jet powers estimated in this way are found to be sufficient to offset cooling, indicating that AGN feedback in the form of relativistic jets is a plausible mechanism of star formation quenching, by depriving the central galaxies of the required cool gas. Observations at radio frequencies reveal that AGN jet feedback may also be important in Milky-Way size galaxies (\citealt{Ledlow2001},\citealt{Singh2015}, \citealt{Nesvadba2021}, \citealt{Webster2021}), as well as dwarf galaxies (\citealt{Pakull2010}, \citealt{Mezcua2019}, \citealt{Yang2020}, \citealt{Davis2022}). Jet feedback may also be relevant in galaxies of various masses at high redshifts ($z>2$; \citealt{Heckman2014}, \citealt{Smolcic2017}).

The theoretical study of jet feedback in massive galaxies has been done largely through hydrodynamical simulations, either zoom-in cosmological simulations (e.g. \citealt{Dubois2010}, \citealt{Bourne2020}), or more commonly in idealised set-ups (e.g. \citealt{Omma2004}, \citealt{Reynolds2006}, \citealt{Yang2019}). In the latter category, a significant effort has been dedicated to studying single jet episodes, either modeling only the hydrodynamical aspect of jets and the bubbles/lobes they inflate (e.g. \citealt{Komissarov1998}, \citealt{Churazov2001}, \citealt{Bruggen2002}, \citealt{Roediger2007}, \citealt{Pavlovski2008}), or including relativistic physics (e.g. \citealt{Walg2013}, \citealt{English2016}, \citealt{Choi2017}), magnetic fields (e.g. \citealt{Hardcastle2014}, \citealt{Tchekhovskoy2016}, \citealt{Mukharjee2020}), radiative cooling (e.g. \citealt{Blondin1990}, \citealt{Stone1997}, \citealt{Guo2018}) or cosmic rays (e.g. \citealt{Guo2011}, \citealt{Ehlert2018}, \citealt{Yang2019}). The main goal of these studies has been to determine the jet energetics, i.e. how much energy is transferred to the ambient medium, where and in what form (e.g. \citealt{Morsony2010}, \citealt{Bourne2017}, \citealt{Weinberger2017}), as well as through which processes (e.g. \citealt{Perucho2010}, \citealt{Bambic}, \citealt{Yang2019}). 

Some simulations in idealised set-ups have also modeled self-consistent accretion, where a central SMBH launches jets based on an accretion rate determined from gas properties near the SMBH. These simulations almost exclusively use adaptive mesh refinement (AMR), with spatial resolutions typically reaching $200-500$ pc (e.g. \citealt{Gaspari2011}, \citealt{Li2015}, \citealt{Beckmann2019}) in the centres of the simulated systems. The jet velocities used are of order $10^4$ $\mathrm{km}\hspace{0.3mm}\mathrm{s}^{-1}$ (e.g. \citealt{Gaspari2011}, \citealt{Yang2016}, \citealt{Meece2017}), and the jet efficiencies $\epsilon_\mr{j}$ (related to the jet power $P_\mr{j}$ and SMBH accretion rate $\dot{M}_\mr{BH}$ through $\epsilon_\mr{j}=P_\mr{j}/\dot{M}_\mr{BH}c^2$) are typically low, in the range $\epsilon_\mr{j}=10^{-4}-10^{-2}$ (e.g. \citealt{Gaspari2012}, \citealt{Yang2016}, \citealt{Martizzi2019}). The jets are usually launched in a fixed direction, but some studies have included precession imposed by hand (e.g. \citealt{Li2017}, \citealt{Meece2017}). The jet powers achieved in these simulations are in the range $P_\mr{j}=10^{45}-10^{46}$ $\mathrm{erg}\hspace{0.3mm}\mathrm{s}^{-1}$ (e.g. \citealt{Yang2016}, \citealt{Li2017}, \citealt{Martin2019}). The cold gas masses found in these simulations are often fairly large, $M_\mr{cold}=10^{10}-10^{11}$ $\mr{M}_\odot$ or larger (e.g. \citealt{Li2014b}), probably due to low jet efficiencies.

In \cite{Husko2022a} we perform hydrodynamical tests of AGN jets with SWIFT, an efficient smoothed particle hydrodynamics (SPH) code (\citealt{Schaller2016}). These tests feature a constant-power jet launched into a constant-density ambient medium. Although AGN jet feedback has been employed in cosmological simulations (e.g. \citealt{Dave2019}), it has not been been tested in such a way with an SPH code, nor has it been resolved to such a degree ($\approx10^6$ particles per jet). We find that the jets and lobes they inflate behave as expected based on the self-similar theory of jet lobe evolution (e.g. \citealt{Kaiser1997}, \citealt{Komissarov1998}), even at very poor resolutions ($\approx500$ particles per jet). These results are relevant for cosmological simulations, as well as simulations with self-consistent jet feedback, where the jets are fed by gas accretion. Such simulations can feature a variety of jet episodes, some of them fairly weak and thus poorly resolved.

In this paper we present results from a study of self-consistent, spin-driven jet feedback in idealised galaxy group and cluster set-ups simulated with SWIFT (see \citealt{Nobels2022} for details of the set-up). Our highest-resolution simulations have a mass resolution of $m_\mr{g}=10^5$ $\mr{M}_\odot$, which is 20 times better than the only other similar SPH simulation of this kind (involving gas cooling, self-consistent jet feedback and star formation) that has been performed (\citealt{Barai2016}). The spatial resolution (gravitational softening length) is $300$ pc in our highest-resolution simulations, matching most of the AMR simulations discussed above. In order to reliably simulate the jet feedback cycle, we also model the evolution of the spin of the central SMBH, including its direction, due to gas accretion and jet spindown. We use high jet efficiencies, based on results of jet launching in general-relativistic magneto-hydrodynamical (hereafter GRMHD) simulations. Modeling SMBH spin evolution results in natural changes in jet direction, as well as emergent jet precession. Similar simulations, involving the modeling of SMBH spin, have recently been performed by \cite{Beckmann2019} using an AMR code. \cite{Luca} also recently studied jet feedback from AGN with an SPH code (as well as other types of AGN feedback), but these were at much higher resolution (pc-scale) and in a different context (disk-type galaxies). We perform simulations in set-ups that span the galaxy group to galaxy cluster regimes (with halo masses from $M_\mr{200}=10^{13}$ $\mr{M}_\odot$ to $M_\mr{200}=10^{15}$ $\mr{M}_\odot$), as well as with varying parameters, in order to probe jet feedback in detail.

In \S~\ref{sec:sec2} we discuss our SMBH spin evolution and jet feedback model. This includes the physics of thick, advection-dominated accretion discs, jet efficiencies from GRMHD simulations, SMBH spinup/spindown from accretion and jets, as well as \cite{LenseThirring} precession. In \S~\ref{sec:sec3} we discuss the numerical implementation of the model, the physical set-up and the different simulations we have done. In \S~\ref{sec:sec4} we lay out the general features of jet feedback, going from the galaxy group scale to the galaxy cluster scale. In \S~\ref{sec:sec5} we discuss jet feedback in more detail using our massive galaxy cluster set-up, focusing on properties of the hot and cold gas. We also present results from variations of feedback-related parameters. In \S~\ref{sec:sec7} we summarise and conclude.

\section{Black hole spin evolution and jet feedback model}
\label{sec:sec2}

The efficiency with which jets are launched from the vicinity of SMBHs depends strongly on the dimensionless spin parameter $a$, which is related to the angular momentum of the SMBH, $J_\mr{BH}$, and its mass, $M_\mr{BH}$, through $a=J_\mr{BH}c/M_\mr{BH}^2G$. We refer to SMBHs with $a=1$ as maximally spinning\footnote{The spin of a SMBH cannot exceed $1$ for theoretical reasons (otherwise the SMBH might feature a naked singularity).}. In the rest of the paper, we assume that $a\in[-1,1]$. Here, positive values represent prograde accretion from the inner accretion disc, whereas negative values represent retrograde accretion, in the case that torques between the inner regions of the disc and the SMBH cause counteralignment (see \S~\ref{sec:prograde_retrograde}).

In the simulations presented in this paper, we include only AGN jet feedback in order to prevent other feedback mechanisms (including stellar and AGN thermal feedback) from interfering with our interpretations of the results. Jets are launched with high efficiencies from SMBHs that accrete slowly, in the thick disc regime (\citealt{NarayanYi1994}). Most supermassive SMBHs that host jets in the local Universe are likely in this accretion regime (\citealt{BestHeckman}, \citealt{Weinberger2017b}). We do not include thin, radiatively-efficient discs (\citealt{ShakuraSunyaev1973}), which are present at high accretion rates, since we do not include radiative (thermal) feedback in our simulations. Below we give a summary of the main properties of thick accretion discs.

\subsection{Thick accretion discs}

Thick accretion discs are known by many names: the advection-dominated accretion flow (ADAF), hot accretion flow, RIAF (radiatively inefficient accretion flow), the hard state (in terms of X-ray spectra) and the low state (in terms of accretion rate). The disc is geometrically thick ($H/R\approx0.5$) and optically thin. The gas in this disc is very hot and diffuse, and thus radiatively inefficient. It is advected inwards and accreted onto the black hole faster than it can radiate away a significant fraction of its thermal energy, resulting in low luminosities. Gas orbits are not fully circular and instead have a significant radial component. The gas flow is continuous all the way down to the event horizon, with no abrupt change in properties at the innermost stable circular orbit ($R_\mathrm{ISCO}$). The poloidal magnetic flux at the event horizon of the SMBH is large, leading to strong jets. We take the solution for this disc from \cite{Narayan1995} (see \cite{YuanNarayan2014} for a detailed review).

The thick accretion disc appears at low (dimensionless) accretion rates of $\dot{m}=\dot{M}_\mathrm{BH,0}/\dot{M}_\mathrm{Edd}\lesssim0.01$, where the Eddington accretion rate is given by
\begin{equation}
\dot{M}_\mr{Edd}=\frac{L_\mr{Edd}}{\epsilon_\mr{r}c^2}=4\pi\frac{G M_\mr{BH}m_\mr{p}}{\epsilon_\mr{r}\sigma_\mr{T}c}.
\label{eq:eq1}
\end{equation}
Here, $m_\mr{p}$ is the proton mass, $\sigma_\mr{T}$ the Thomson cross-section and $\epsilon_\mr{r}=L_\mr{bol}/\dot{M}_\mr{BH}c^2=0.1$ a nominal radiative efficiency used only for the definition of $\dot{m}$ in this paper (we do not include radiative feedback, nor is the radiative efficiency as high as $0.1$ for the thick disc). $\dot{M}_\mathrm{BH,0}$ is the large-scale accretion rate of the SMBH (before the matter settles down to an accretion disc).

\subsection{Jet efficiency}
\label{sec:jet_eff}

According to the model of \cite{Blandford1977} (BZ), magnetic fields present due to an accretion disc plunge into the SMBH's ergosphere and corotate due to frame dragging, resulting in a net outward flux of energy and angular momentum. The power of the jet that is launched in the BZ process scales as $P_\mr{jet}\propto\Omega_\mr{H}^2\Phi_\mr{H}^2$, where $\Omega_\mr{H}$ is the angular velocity of the event horizon, and $\Phi_\mr{H}$ is the net poloidal magnetic flux threading the horizon. The largest source of uncertainty in modeling jet powers comes from the strength of the magnetic field, which determines the flux $\Phi_\mr{H}$. GRMHD simulations of thick discs find that they settle down to the equilibrium magnetically-arrested disc state (MAD, \citealt{Narayan2003}). The large poloidal magnetic field in the central regions of the disc `chokes` the inward flow, causing the accretion to proceed in discrete blobs (or thin streams at very high resolution, see \citealt{Ripperda2022}). Simulations of the jet launching process in these systems have converged in terms of how much energy the jets extract from the SMBH (e.g. \citealt{Tchekhovskoy2011}, \citealt{McKinney2012}, \citealt{Sadowski2014}, \citealt{Liska2020}, \citealt{Narayan2021}).

Observational inferences indicate that most thick discs are in the MAD state (\citealt{Ghisellini2014}). Recent direct measurements of the magnetic field in the thick disc surrounding the central SMBH in M87 confirm this (\citealt{EHT2021}). High-resolution and long-duration simulations have found that the MAD state is achieved even without any initial poloidal magnetic field, bolstering the theoretical expectation that all thick discs should be MAD (\citealt{Liska2020}). Simulations of thinner accretion discs have also found that the MAD state can be achieved in those systems (\citealt{Liska2019}), and the jet powers are then much higher than classically expected (\citealt{Meier2002}).

\begin{figure}
\includegraphics[width=1.01\columnwidth, trim = 0 10 0 0]{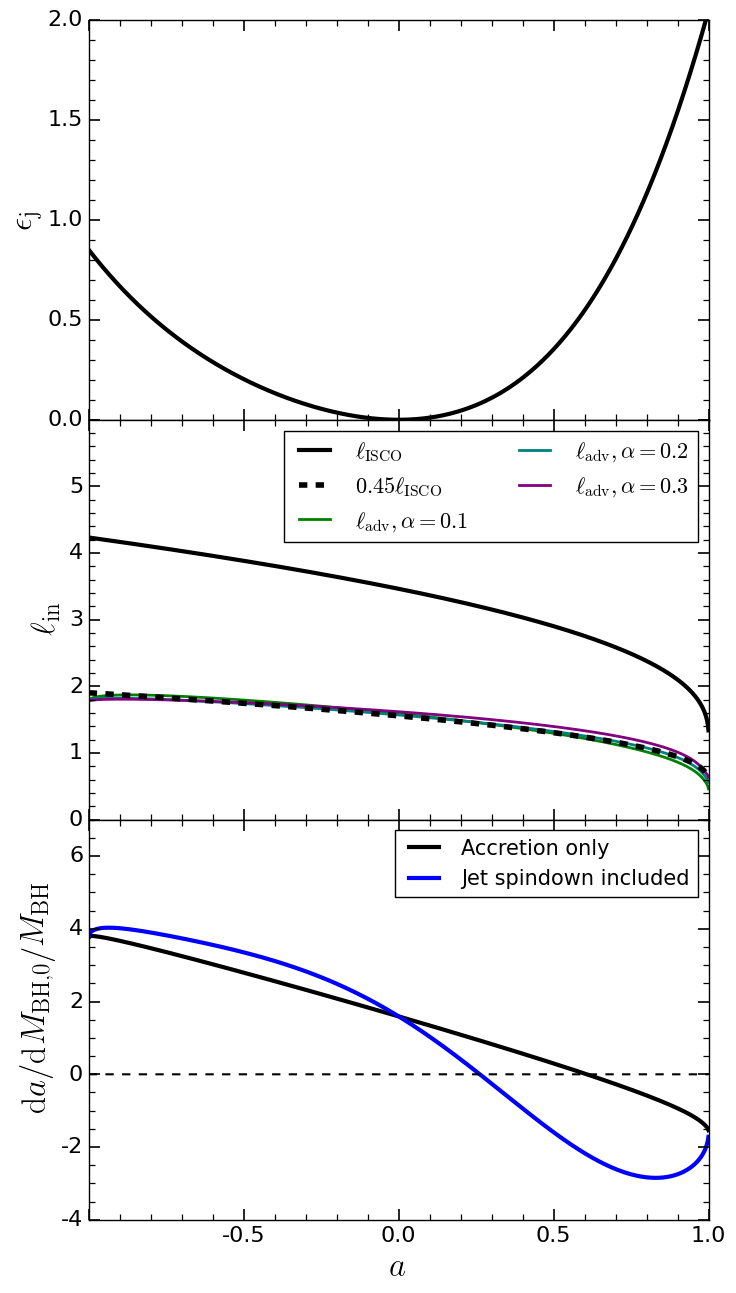}
\caption{The dependencies on spin of the jet efficiency (top), (dimensionless) specific angular momentum at the inner radius (middle) and spinup/spindown function for thick, advection-dominated discs (bottom). The jet efficiency is given by equation (\ref{eq:epsilon_jet}). In the middle panel, coloured lines show the fitting function for the specific angular momentum of accreting matter from \protect\cite{Benson2009}, for a few possible values of accretion disc viscosity $\alpha$. These are all approximately consistent with $45$ per cent of the thin-disc specific angular momentum at the ISCO (innermost stable circular orbit, see Appendix \ref{sec:app1} for expression), for all spins. In the bottom panel, the black line (equation \ref{eq:da_dlnMSMBH}) shows that SMBHs surrounded by thick discs never spin up beyond $a=0.6$, while the blue line (adding the equation \ref{eq:da_dlnMSMBH_jet} term) shows that jets bring this equilibrium spin value down to $0.25$.}
\label{fig:fig0}
\end{figure}

The jet power in the MAD state is proportional to the accretion rate, and the relation is usually expressed in terms of the jet efficiency $\epsilon_\mr{j}$ as $P_\mr{j}=\epsilon_\mr{j}\dot{M}_\mr{BH,0}c^2$. We use the spin-dependent jet efficiency formula found by \cite{Tchekhovskoy2010}, which is based on GRMHD simulations and is applicable for thick accretion discs in the MAD state. The jet efficiency is given by
\begin{equation}
    \epsilon_\mr{j}=\frac{\kappa}{4\pi}\phi_\mr{BH}^2\Omega_\mr{BH}^2\big[1+1.38\Omega_\mr{BH}^2-9.2\Omega_\mr{BH}^4\big],
\label{eq:epsilon_jet}
\end{equation}
where $\kappa$ is a numerical factor that depends on the initial geometry of the magnetic field (e.g. 0.054 for split-monopole vs. 0.044 for parabolic, we assume $\kappa=0.05$), $\phi_\mr{BH}$ is the dimensionless magnetic flux threading the horizon (see \citealt{Tchekhovskoy2010} for the precise definition), and $\Omega_\mr{BH}=a/2r_\mr{H}$ is the (dimensionless) angular velocity of the SMBH event horizon. Here, $r_\mr{H}=1+\sqrt{1-a^2}$ is the radius of the horizon in units of the gravitational radius $R_\mr{G}=M_\mr{BH}G/c^2$. Equation (\ref{eq:epsilon_jet}) agrees very well with the results from higher-resolution simulations performed by \cite{Narayan2021}, who provide the following fit for the magnetic flux as a function of spin:
\begin{equation}
    \phi_\mr{BH}(a)=-20.2a^3-14.9a^2+34a+52.6.
\label{eq:phi_a}
\end{equation}
The jet efficiency given by equation (\ref{eq:epsilon_jet}) has a strong dependence on spin; for low values of spin it scales as $\epsilon_\mr{j}\sim a^2$, whereas for high values the dependence is even stronger ($\epsilon_\mr{j}\sim a^4-a^6$). In the top panel of Fig. \ref{fig:fig0} we show the dependence of the jet efficiency on spin. The difference between prograde and retrograde accretion is clearly visible. At $a=0.5$, the jet efficiency is $\approx30$ per cent, while at $a=1$, it is around $200$ per cent (indicating a net decrease in the total mass-energy). Retrograde SMBHs never launch jets with efficiencies above $100$ per cent.

\subsection{Accretion spinup/spindown}

The primary mechanism of SMBH spin evolution is accretion of matter, facilitated by the existence of an accretion disc. The change in the magnitude of the angular momentum of the SMBH can be related to the accretion rate through the relation 
\begin{equation}
    \frac{\mr{d}J_\mr{BH}}{\mr{d}t}=L_\mr{in}\frac{\mr{d}M_\mr{BH,0}}{\mr{d}t},
\label{eq:dJ_dt}
\end{equation}
where $L_\mr{in}$ is the specific angular momentum of accreting matter at some inner radius $R_\mr{in}$, and we have ignored jet spindown (for now). $R_\mr{in}$ is the radius within which matter does not efficiently transport angular momentum or energy outwards. Equation (\ref{eq:dJ_dt}) can be translated into an equivalent equation for spin evolution (\citealt{Bardeen1970}, \citealt{Fanidakis2011})\footnote{Note that time does not appear as an independent variable in the equation for spin evolution. Instead, the change in spin is determined entirely by the the current value of spin and the amount of matter being accreted.}:
\begin{equation}
    \bigg(\frac{\mr{d}a}{\mr{d}M_\mr{BH,0}/M_\mr{BH}}\bigg)_\mathrm{acc}=\ell_\mr{in}-2a e_\mr{in},
\label{eq:da_dlnMSMBH}
\end{equation}
where $\ell_\mr{in}=cL_\mr{in}/GM_\mr{BH}$ is the dimensionless specific angular momentum. The change in mass of the SMBH can be related to the mass funneled towards it from large distances through $\mr{d}M_\mr{BH}=(1-e_\mr{r})\mr{d}M_\mr{BH,0}$. The second term in equation (\ref{eq:da_dlnMSMBH}) originates from the definition of spin, $a=J_\mr{BH}c/M_\mr{BH}^2G$, which results in two terms if a derivative is taken. The same term includes the specific binding energy $e_\mr{in}$, which we assume to be $e_\mr{in}=1$ (see \citealt{Benson2009} for the effects of varying the choice of $e_\mr{in}$). This corresponds to the assumption that the radiative efficiency is negligible in the thick disc (see e.g. \citealt{Mahadevan} or \citealt{YuanNarayan2014}), and also that the transport of energy outwards through viscous or magnetic forces is negligible.

Orbits in the thick disc are not circular and stable out to some radius $R_\mr{in}$; gas properties instead vary with radius smoothly down to the event horizon of the SMBH, so that $R_\mr{in}=R_\mr{H}=R_\mr{G}(1+\sqrt{1-a^2})$. Self-similar solutions for the thick disc (e.g. \citealt{NarayanYi1994}) assume Newtonian gravity, which means that they are only correct at large distances, typically $r>10R_\mr{G}$. We instead take the values for $\ell_\mr{in}$ at the event horizon based on numerical calculations done by \cite{PophamGammie1998}, who studied advection-dominated accretion flows in the Kerr metric (\citealt{Kerr}) for various values of spin $a$, adiabatic index $\gamma$, advection parameter $f$ (see e.g. \citealt{YuanNarayan2014} for definition) and viscosity parameter $\alpha=\nu/c_\mr{s}H$, where $\nu$ is the kinematic viscosity, $c_\mr{s}$ the sound speed and $H$ the thickness of the disc. In particular, we take the fitting function found by \cite{Benson2009}, which represents these results quantitatively. We assume purely advection-dominated flows ($f=1$).

In the middle panel of Fig. \ref{fig:fig0} we show the specific angular momentum from \cite{Benson2009} for a few values of $\alpha$, showing that the dependence on $\alpha$ is very weak. We also show the specific angular momentum at the innermost stable circular orbit (ISCO, see Appendix \ref{sec:app1} for the expression), assuming fully circular orbits. This is appropriate for the thin disc (\citealt{ShakuraSunyaev1973}, \citealt{NovikovThorne1973}). The dashed line shows a scaled-down ISCO specific angular momentum. According to the \cite{Benson2009} fitting function, $\ell_\mr{in}$ is roughly $45\%$ that of the thin disc value for all values of spin. For simplicity, we assume $\ell_\mr{in}=0.45\ell_\mr{ISCO}$ for the remainder of this paper. This finding for the value of $\ell_\mr{in}$ is similar to that from Newtonian self-similar models. For the thick disc, equations from \cite{Narayan1995} imply an orbital velocity that is $0.25-0.37$ of the Keplerian one for $\alpha=0.3-0.05$, which is close to the correct general-relativistic value. 

The value of $\alpha$ that we use is based on numerical results and observations (note that numerical simulations give only a value for the product $\alpha \beta$, $\beta$ being related to the magnetic-to-total pressure ratio). Numerical results indicate that hot accretion flows (thick accretion discs) appear, in one form or another, for dimensionless accretion rates $\dot{m}<0.4\alpha^2$ (\citealt{YuanNarayan2014}). Observational studies based on analysing AGN spectra find that the transition from a thick to a thin disc occurs at $\dot{m}=0.02-0.03$ (e.g. \citealt{Russell2013}, \citealt{Noda2018}). Combining this finding with the numerical results, $\alpha$ can be constrained to the range $0.2-0.3$. This is in agreement with more direct observational estimates that also find $\alpha=0.2-0.3$ (e.g. \citealt{Martin2019}). In this paper we assume $\alpha=0.2$ (note that this is twice as large as is often assumed, e.g. \citealt{Griffin2019a}).

In the bottom panel of Fig. \ref{fig:fig0} we show the spinup/spindown function in the case that only gas accretion is included, given by equation (\ref{eq:da_dlnMSMBH}), with the assumed $\ell_\mr{in}$ for the thick disc. This shows that the SMBH will spin up if accretion is retrograde ($a<0$), and also if it is prograde ($a>0$) and that spin is $a\lesssim0.6$. If the spin is larger than that, the SMBH will spin down. This is somewhat confusing$-$how can pure accretion lead to SMBH spindown? The answer lies in a combination of frame-dragging and viscous stresses; some of the angular momentum of the SMBH is transferred to the gas orbiting around it. These particles are on fairly radial orbits in the thick disc, and frame-dragging can accelerate their orbital velocities on account of the spin of the SMBH. In this process, some of the angular momentum is transferred outwards through viscous forces, resulting in spindown.


\subsection{Jet spindown}

The effects of jets on SMBH spin evolution can be encapsulated as an additional term to be added to equation (\ref{eq:da_dlnMSMBH}), which can be written as (see \citealt{Benson2009} for derivation):
\begin{equation}
    \bigg(\frac{\mr{d}a}{\mr{d} M_\mr{BH,0}/M_\mr{BH}}\bigg)_\mr{j}=-\epsilon_\mr{j}(a)\frac{\sqrt{1-a^2}}{a}\bigg[\Big(\sqrt{1-a^2}+1 \Big)^2+a^2 \bigg].
\label{eq:da_dlnMSMBH_jet}
\end{equation}
Here we have ignored the effects of disc winds (unlike \citealt{Benson2009}), which would generally appear as an additional efficiency term along with $\epsilon_\mr{j}$. The derivation of equation (\ref{eq:da_dlnMSMBH_jet}) assumes that the launching of the jet and accretion are decoupled processes, i.e. the mass-energy of the gas in the accretion disc does not directly contribute to the jet, and it is instead powered entirely by the rotational energy of the SMBH.

A further assumption in the derivation of equation (\ref{eq:da_dlnMSMBH_jet}) is that the change of rotational energy of the SMBH, $\dot{E}_\mr{rot}$, exactly matches the jet power (in magnitude). This is equivalent to assuming that the irreducible mass-energy of the SMBH, $E_\mathrm{irr}$ (which is related to the rotational energy through $E_\mathrm{rot}+E_\mathrm{irr}=c^2M_\mathrm{BH}$), remains constant as the jet is launched. While the irreducible mass-energy cannot be reduced in the jet launching process, it is possible that the irreducible mass-energy grows as the jet is launched, with the rotational mass-energy being decreased at a rate even greater (in magnitude) than $-P_\mathrm{j}$. Thus, equation (\ref{eq:da_dlnMSMBH_jet}) represents a \textit{minimum} spindown rate due to jet launching.

Equation (\ref{eq:da_dlnMSMBH_jet}) shows that stronger jets spin down the SMBH more than pure accretion, as expected. Simulations of jet launching in the MAD state find that jets spin down the SMBH very effectively (e.g. \citealt{Narayan2021}), showing that the jet spindown term cannot be ignored in the evolution of SMBH spin (as has often been assumed). From the bottom panel of Fig. \ref{fig:fig0}, we see that including jet spindown in equation (\ref{eq:da_dlnMSMBH}) results in faster spindown for retrograde spins (note that positive values of $s$ for negative values of $a$ indicate spindown), and a lower equilibrium spin value for prograde accretion ($0.25$ instead of $0.6$ without jets).

Note that simulations of MAD jets imply stronger spindown than we have assumed here, to the point of the equilibrium spin value being $a\approx0$ (\citealt{Narayan2021}). In an idealised set-up with only thick discs (such as the one we are presenting here), an equilibrium spin value of $\approx0$ would imply an effectively finite amount of energy that an accreting SMBH can launch in the form of jets, before being spun down to $a\approx0$. This would be problematic in our idealised simulations, since the SMBHs would cease to do any feedback once they are spun down. As a result, we do not implement the spindown-related findings from MAD simulations in this paper, and we instead use the analytical prescription above. Note that in a more realistic scenario, including thin, radiatively efficient discs at high accretio rates, as well as SMBH mergers, equilibrium values of $a\approx0$ are not problematic. This is because the SMBH accretion rate would simply increase, as a result of a lack of jet feedback, until the SMBH enters the thin disc regime, where it can more effectively spin up.

\subsection{The structure of the disc: bending wave regime}
\label{sec:structure}

We have so far discussed how the magnitude of spin evolves given the current spin and the mass accretion rate. However, we have not stated what we assume for the direction of the spin. The spindown of SMBHs due to jet launching only changes the magnitude of the spin, and not its direction. For accretion, we could assume that the angular momentum change, corresponding to equation (\ref{eq:da_dlnMSMBH}), is in the direction of the large-scale angular momentum of the gas surrounding the SMBH. We measure this direction using the SPH smoothing kernel around the SMBH in the simulation, which we denote by $\mathbf{\hat{J}_\mr{d}}$. A complication to this procedure, however, is that the SMBHs also experience \cite{LenseThirring} torques, which we discuss below. In addition, the specific angular momentum of accreting matter, as well as the feedback efficiencies, depend critically on whether accretion is prograde ($a>0$) or retrograde ($a<0$). In order to determine this, more detailed accretion disc physics must be included in our model.

Spinning SMBHs induce \cite{LenseThirring} precession  (hereafter LT) of a parcel of gas orbiting the SMBH, as a result of torques that are related to the frame-dragging of space-time in the \cite{Kerr} metric. In the context of an accretion disc, the LT torque can have different effects depending on which accretion regime the disc is in (see \citealt{Nixon2016} for a review). In all cases, LT precession is effective only within some radius $R_\mr{warp}$. Within that radius, torques between the disc and the SMBH effectively facilitate the transfer of angular momentum between the two, whereas outside it no such transfer occurs.

The effects of LT precession depend on the ratio of the viscosity parameter of the disc, $\alpha$, and its aspect ratio $H/R$. In the case $\alpha\gg H/R$ (thin disc), the disc is aligned or counter-aligned with the SMBH spin vector out to the radius $R_\mathrm{warp}$ (\citealt{Papaloizou1983}, \citealt{BardeenPetterson}), and thus has a warped shape (hence the name). In the case of a thick disc $\alpha\ll H/R$, so-called bending waves cause the precession of the disc within some inner radius, with the precession rate depending on radius (\citealt{Ogilvie1999}, \citealt{Lubow2002}, \citealt{King2005}). 


\cite{Lubow2002} found that the behaviour of precessing discs in the bending wave-regime depends on the value of a dimensionless variable $x$ given by
\begin{equation}
x=\left(\frac{24 \vert a \vert}{h^2}\right)^{1 / 2} \frac{r^{-(p+1 / 4)}}{p+\frac{1}{4}},
\label{eq:x}
\end{equation}
where $h$ and $p$ are used to parametrize the aspect ratio as $H/R=h r^{p-1}$. The aspect ratio in the thick disc does not depend on radius and is equal to some value $H/R$. In terms of $h$ and $p$, this choice corresponds to $p=1$ and $h=H/R$. For $x\ll1$ (large $R$), the disc is unaffected by the bending waves and remains aligned with the large-scale direction of angular momentum $\mathbf{\hat{J}_\mr{d}}$. In the inner regions ($x\gg 1$), it experiences precession. The transitional radius between the two regions can be found by taking $x=1$ and inverting equation (\ref{eq:x}), this yields
\begin{equation}
R_\mr{warp,adv}=R_\mr{G}\bigg(\frac{384\vert a\vert}{25(H/R)^2}\bigg)^{2/5}.
\label{eq:r_warp_adaf}
\end{equation}
For the thick disc, we take $H/R=0.3$ based on GRMHD simulations (note that this is less than the value in the self-similar solution of \citealt{Narayan1995}), resulting in values of $R_\mr{warp}$ that are at most $R_\mr{warp,adv}\approx6R_\mr{G}$. In the thin disc, instead of causing precession, the LT torques cause the disc to be perfectly aligned or counter-aligned out to thousands of $R_\mathrm{G}$, due to the \cite{BardeenPetterson} effect. We find similar values for the warp radius, of order several $R_\mr{G}$, if we instead assume that the Bardeen-Peterson effect operates for the thick disc.

In deriving equation (\ref{eq:x}), \cite{Lubow2002} considered the case of a nearly Keplerian disc with a weak tilt. Despite these assumptions, observations have found that bending wave radii given by equation (\ref{eq:r_warp_adaf}) are able to explain quasi-periodic oscillations in light curves of X-ray binaries thought to host hot accretion flows (\citealt{Ingram2009}, \citealt{Ingram2012}). In addition, simulations have reproduced the findings from \cite{Lubow2002} (e.g. \citealt{Fragile2007}, \citealt{Fragile2009} \citealt{Liska2018}). 


\subsection{Prograde and retrograde accretion}
\label{sec:prograde_retrograde}

In our model, we decide whether accretion is prograde or retrograde in the same way as in \cite{Griffin2019a} (see also \citealt{Volonteri2007}, \citealt{King2008}, \citealt{Fanidakis2011}). The relevant quantities in this algorithm are the SMBH angular momentum $\bf{J_\mr{BH}}$ and the disc angular momentum within the region of influence of LT precession. We take the latter as the angular momentum within the warp radius, $\bf{J_\mr{warp}}$. The actual angular momentum of the entire accretion disc is, of course, very different. However, the angular momentum outside $R_\mr{warp}$ is irrelevant for this purpose since those regions of the disc do not interact with the SMBH through LT torques. 

We assume that the direction of the angular momentum of the accretion disc on large scales (outside the warp radius) matches that which we measure around the SMBH in our simulations (using SPH particles in the SMBH smoothing kernel). This is a fairly strong assumption, given the fact that we measure the angular momentum direction on scales of $100-1000$ pc, while the subgrid accretion disc is of order $1$ pc or smaller. We typically resolve the Bondi radius, at least in our high-resolution simulations, so the direction of angular momentum being funneled towards the black hole should remain the same even on unresolved scales, if the black hole accretes directly from the hot phase. This assumption may break down in the chaotic cold accretion scenario proposed by \cite{Gaspari2013}. Related to this issue is the possibility that accretion on small scales may be inherently chaotic (with effectively random directions), due to turbulence in the cold, star-forming interstellar medium around the black hole (\citealt{King2006}, \citealt{Hopkins2012}). This is supported by some observations (e.g. \citealt{Greene2013}, \citealt{Ruffa2020}), but not by others (e.g. \citealt{Kaviraj2015}, see also tentative evidence from the Event Horizon Telescope observations that suggests misalignment of $<30\degree$ between the Sagittarius A* accretion disc and the Milky Way gas disc; \citealt{EHTSagA5}). This issue, if it does exist, can only be overcome by resolving the gas around the SMBH on much smaller scales (\citealt{Angles2021}), which is far beyond the reach of current cosmological simulations. Until this becomes possible, our assumption allows us to model the spins of black holes, if not fully realistically.

In our model, accretion proceeds in finite increments, such that one warp mass $M_\mr{warp}$ is swallowed at a time (with $M_\mr{warp}$ the disc mass within $R_\mr{warp}$). Before $M_\mr{warp}$ is swallowed, the torques between the SMBH and the disc bring the system to a steady state. During this process, the magnitude of the SMBH angular momentum remains constant, while its direction gradually aligns with that of the total angular momentum $\bf{J_\mr{tot}}=\bf{J_\mr{BH}}+\bf{J_\mr{warp}}$. The angle between $\bf{J_\mr{BH}}$ and $\bf{J_\mr{warp}}$ decreases with time, also resulting in the decrease in the magnitude of $\bf{J_\mr{warp}}$. 

Our procedure effectively amounts to assuming that torques between the SMBH and the inner disc first reorient the SMBH, before any matter accretes. This allows us to assume that the accreting matter is either aligned or counteraligned with respect to the new spin axis (prograde or retrograde accretion, respectively). Accretion is retrograde if
\begin{equation}
\cos \theta<-\frac{J_{\mathrm{warp}}}{2 J_{\mathrm{BH}}},
\label{eq:counteralignment}
\end{equation}
where $\cos \theta=\bf{\hat{J}_\mr{BH}}\cdot\bf{\hat{J}_\mr{d}}$ is the initial misalignment between the SMBH and the (large-scale) angular momentum of the disc, whose direction is $\bf{\hat{J}_\mr{d}}$ (see \citealt{King2005} for a derivation). In the case that equation (\ref{eq:counteralignment}) is not satisfied, accretion is assumed to be prograde. Note that for the thick disc, the assumption of (counter-)alignment is not strictly correct; the disc actually precesses. However, the precession is assumed to proceed around the new axis, so that \textit{on average} the thick disc is also (counter-)aligned. 

The warp angular momentum in equation (\ref{eq:counteralignment}) is calculated by integrating the product of the surface density of the thick disc and $L(R)$, the specific angular momentum at a distance $R$ from the SMBH, out to $R_\mr{warp}$. A similar integral (without the $L(R)$ factor) is used to calculate the warp mass $M_\mr{warp}$. We use the surface density from the self-similar thick disc solution presented in \cite{Narayan1995}:
\begin{equation}
\Sigma_\mr{adv}=\frac{\dot{M}_\mathrm{BH,0}}{2\pi R\vert v_\mr{r} \vert},
\label{eq:Sigma_adv}
\end{equation}
where $v_\mr{r}=-\alpha v_0 v_\mr{K}$ is the radial velocity. Here, $v_\mr{K}=\sqrt{M_\mr{BH}G/R}$ is the Keplerian velocity, and $v_0$ is a numerical coefficient. The specific angular momentum is given by $L(R)=\Omega_0\sqrt{M_\mr{BH}GR}$, where $\Omega_0$ is another numerical coefficient. The two numerical coefficients are calculated as $v_0=3/(5+2\varepsilon)$ and $\Omega_0=\sqrt{2\varepsilon/(5+2\varepsilon)}$, where $\varepsilon=(5/3-\gamma)/(\gamma-1)$. The adiabatic index $\gamma$ can be related to the gas-to-total pressure ratio $\beta$ (\citealt{Esin}):
\begin{equation}
    \gamma = \frac{8-3\beta}{6-3\beta}.
\label{eq:gamma_beta}
\end{equation}
Finally, we connect $\beta$ to $\alpha$ using findings from GRMHD simulations: $\beta=1/(1+2\alpha)$ (\citealt{YuanNarayan2014}). $v_0$ varies weakly with $\alpha$; for $\alpha=0.05$, it is $0.56$, whereas for $\alpha=0.2$, it evaluates to $0.52$. $\Omega_0$ depends on $\alpha$ somewhat more strongly; we obtain $0.27$ and $0.37$ for the same values of $\alpha$. 


\section{Numerical implementation and physical set-up}
\label{sec:sec3}

\subsection{Numerical code and subgrid physics}
\label{sec:SWIFT}

We use SWIFT (\citealt{Schaller2016}), an open-access\footnote{\href{https://swiftsim.com}{https://swiftsim.com}} simulation code that includes hydrodynamics, gravity, cosmology, as well as various subgrid physical processes. This includes our model for the evolution of BH spin, which is available to use as part of the code. SWIFT is currently being used in large simulations such as the SIBELIUS suite (\citealt{McAlpine2022}), and will be used in upcoming successors to the EAGLE simulation (\citealt{Schaye2015}). It is a Lagragian code based on smoothed particle hydrodynamics (SPH; \citealt{Monaghan1992}). We use the SPHENIX hydrodynamical implementation in SWIFT (\citealt{Borrow2022}), which includes artificial viscosity and conduction (as well as respective limiters). Both are necessary in order to solve the hydrodynamics equations in the general sense, but they are particularly important when attempting to simulate extreme contrasts in fluid properties, such as those present in supernova and AGN feedback events.




In our simulations, we represent the gravity of the dark matter halo using an external potential. The stellar component is represented by a live population of gravitationally interacting particles, while the gaseous component is represented with SPH particles. The smoothing lengths are set to 1.2348 times the local inter-particle seperation, corresponding to a target neighbour number of 58. The minimal smoothing lengths are set to 0.01 times the gravitational softening length (the values of which are discussed in \S \ref{sec:sims}).

The gas is allowed to cool radiatively based on the cooling function from \cite{Ploeckinger2020}, although it is not allowed to cool down to the molecular phase. Instead, we use an entropy floor (see \citealt{Nobels2022} for details). Star formation is modeled based on the Kennicutt-Schmidt law (\citealt{Kennicutt}) using the gas pressure (\citealt{Schaye2008}). We do not include any stellar feedback, magnetic fields or other physics.

\subsection{Black hole accretion}
\label{sec:bh_accretion}

In the centre of the halo we place a SMBH and fix its position, not allowing it to wander around based on gravitational interactions with the surrounding gas, nor to reposition to the potential minimum (\citealt{Bahe2022}). We model the accretion rate using the Bondi-Hoyle-Lyttleton prescription (\citealt{Hoyle}, \citealt{Bondi}):
\begin{equation}
\dot{M}_\mr{B}=4\pi\frac{G^2M_\mr{BH}^2\rho}{(c_\mr{s}^2+v^2)^{3/2}},
\label{eq:bondi}
\end{equation}
where $\rho$, $c_\mr{s}$ and $v$ are the kernel-weighted density, isothermal sound speed and velocity (relative to the SMBH) of the gas, respectively, all of which are calculated from the smoothing kernel of the SMBH. We assume that $\dot{M}_\mr{BH,0}$, the large-scale accretion rate of the SMBH, is equal to the Bondi rate. Here we use the subscript '0' to differentiate the large-scale accretion rate and the mass growth rate $\dot{M}_\mr{BH}$; the two differ since the radiative and/or jet efficiencies are non-zero.

Some observations (e.g. \citealt{Nemmen2015}) indicate that a fraction (possibly a very large fraction) of the material infalling from the Bondi radius does not reach the black hole. Instead, it could be blown away in a kinetic wind (\citealt{Blandford1999}, \citealt{YuanNarayan2014}), effectively reducing the feedback efficiency. Most simulations with a similar set-up as ours have used low efficiencies, since such efficiencies appear to be in line with observations. For simplicity, and since we are presenting the first application of a model with self-consistent, spin-driven jet feedback hosted by a thick accretion disk, we do not reduce the Bondi accretion rate by any such factor. For a similar reason, we do not suppress the Bondi rate due to the turbulence and vorticity of the gas (e.g. \citealt{Krumholz2005}, \citealt{Krumholz2006}). Our results should thus be treated as an upper limit to the possible impact of jets.


\subsection{The numerical algorithm for spin evolution}
\label{sec:num_algorithm}

In the previous section we discussed the theory behind our model for spin evolution. Here we will lay out how we implement the model, and and how this can be generalized to other hydrodynamical simulations (e.g. EAGLE, \citealt{Schaye2015}), and in general in any hydrodynamical code (e.g. SWIFT, \citealt{Schaller2018}). Using the same SPH particles that are used to measure the Bondi accretion rate onto the SMBH, we measure the angular momentum direction of the gas, $\bf{\hat{J}_\mr{d}}$, in SMBH smoothing kernel. We assume this to be the direction of the angular momentum of the subgrid accretion disc at large distances, outside the influence of LT torques (i.e. outside the warp radius).


At the beginning of every time step of length $\Delta t$, given a mass reservoir $\Delta M_0=\dot{M}_\mr{BH,0}\Delta t$ to be consumed and disc angular momentum direction $\bf{\hat{J}_\mr{d}}$, our algorithm for evolving SMBH-related quantities is as follows:
\begin{enumerate}
    \item Calculate the warp radius $R_\mr{warp}$, mass $M_\mr{warp}$ and angular momentum $J_\mr{warp}$ (\S~\ref{sec:structure}).
    \item Decide whether accretion is prograde or retrograde, based on the angle between the current SMBH angular momentum direction $\bf{\hat{J}_\mr{BH}}$ and that of the disc $\bf{\hat{J}_\mr{d}}$, as well as the ratio of warp and SMBH angular momenta (equation \ref{eq:counteralignment}). If prograde, we set $a=+\vert a\vert$, and if retrograde $a=-\vert a\vert$.
    \item Calculate the jet feedback efficiency $\epsilon_\mr{j}$ (\S~\ref{sec:jet_eff}).
    \item Increase the SMBH mass by $(1-\epsilon_\mr{j})\Delta M_0$ and evolve the SMBH spin according to equations (\ref{eq:da_dlnMSMBH}) and (\ref{eq:da_dlnMSMBH_jet}), i.e. including spinup/spindown from accretion, as well as the term responsible for jet spindown. The direction of the angular momentum of the SMBH is modified such that it matches that of $\mathbf{J_\mr{BH}}+N_\mathrm{warp}J_\mr{warp}\mathbf{\hat{J}_\mr{d}}$, where $\bf{J_\mr{BH}}$ is the old SMBH angular momentum vector, and $N_\mr{warp}$ is defined below. The jet feedback energy reservoir is incremented by $\epsilon_\mr{j}c^2\Delta M_0$.
 \end{enumerate}
Note that step i) can precede step ii) since warp-related quantities do not depend on the sign of $a$. In the above algorithm, $N_\mathrm{warp}=\Delta M_0/M_\mr{warp}$ represents the number of individual accretion events assumed to occur over a single time step. This can also be viewed as the SMBH acquiring angular momentum through LT torques from the warped disk with an effective specific angular momentum of $L_\mathrm{warp}=J_\mathrm{warp}/M_\mathrm{warp}$, so the total angular momentum acquired by the SMBH is $\Delta J =L_\mathrm{warp}\Delta M_0 = (J_\mathrm{warp}/M_\mathrm{warp}) \Delta M_0 = N_\mathrm{warp} J_\mathrm{warp}$. Note that typically, $N_\mathrm{warp}\gg1$, due to small warp radii of the thick disc, and thus also small warp masses. Numerically, it is not feasible to evolve the system one warp increment at a time (nor is there any gain in doing so). Finally, the above algorithm is only applicable if the black hole spin and its direction change very little over a single time step. We ensure this by adding a SMBH time-step whose duration is chosen such that $\Delta J\approx0.01J_\mathrm{BH}$. 

In the Appendix \ref{sec:app2} we show that the timescale for alignment of the SMBH spin vector with that on large scales using this scheme is similar to the alignment timescale in an approach where LT torques are explicitly included in the equation for angular momentum evolution. This demonstration was done for the thin, radiatively-efficient disc (\citealt{ShakuraSunyaev1973}), rather than the thick disc, since the relevant LT torque terms in the angular momentum evolution equation are valid only for the thin disc. We found that the timescale in our warp increment approach is $\approx10\%$ longer, but it depends on SMBH mass and accretion rate the same way as the one in the differential equation approach. 

\subsection{Jet launching}
\label{sec:jet_launching}

The jet power is calculated from the current spin and mass accretion rate as $P_\mr{j}=\epsilon_\mr{j}\dot{M}_\mr{BH,0}c^2$, using the spin-dependent efficiency presented in \S~\ref{sec:jet_eff}. With every time step $\Delta t$, the jet energy reservoir is increased by $P_\mr{j} \Delta t $. When this reservoir exceeds $2\times(1/2)m_\mr{g}v_\mr{j}^2$, where $m_\mr{g}$ is the particle mass and $v_\mr{j}$ the launching velocity, two particles are kicked from the SMBH smoothing kernel,\footnote{Note that energy is not exactly conserved with this scheme. However, since our launching velocities are always much larger than the initial ones, this effect is negligible.} and the jet reservoir is decremented by $2\times(1/2)m_\mr{g}v_\mr{j}^2$. The two particles kicked in each jet event are the farthest from the SMBH in our standard scenario, with one on each side of the SMBH (relative to its angular momentum vector). We choose the farthest particles as our fiducial prescription since we found that other choices can lead to rapid evacuation of the region around the SMBH. The velocity vectors are chosen at random within cones with half-opening angles, relative to the spin axis, equal to some value $\theta_\mr{j}$ (our standard choice being $10\degree$). We compute jet powers using actual jet kicking events (with adaptive time bins, each with a target number of 20 kicking events), instead of defining it as the rate at which the the jet reservoir is increased due to accretion.

The jet launching velocity, $v_\mr{j}$, is a free parameter in our model, and probably the most important one (see \citealt{Husko2022a}). Choosing values that are too low leads to high-momentum (ballistic) jets that drill through the gaseous halo, without experiencing significant shocks, inflating bubbles or heating the gaseous halo. Real AGN jets are highly relativistic and low-density, thus reaching the self-similar stage very quickly, or equivalently at very small distances (see e.g. \citealt{Kaiser2007} for the physics of jets in the self-similar regime). On the other hand, using very large values of the jet launching velocity (close to relativistic) leads to poorly resolved jets. Note that the evolution of the shapes of the self-similar lobes inflated by jets in the self-similar regime, as well as their energetics, should not vary at all with velocity (\citealt{Kaiser2007}). Furthermore, non-relativistic jets that inflate self-similar lobes are very similar to self-similar lobes produced by relativistic jets of the same jet power, with differences of order $10\%$. Through trial and error we have found that values of $v_\mr{j}\approx10-30c_\mathrm{s}$, where $c_\mathrm{s}$ is the sound speed of the ICM, represent a reasonable compromise. 

The choice of an appropriate velocity can ensure that the jet-inflated lobes in our simulations reach the self-similar regime. This in turn means that they exhibit similar hydrodynamics as in the case where they are inflated by fully relativistic jets. However, it is important to note that such jet-inflated lobes do not capture all aspects of observed radio lobes. This is because we do not model physics that may be important for this particular problem. Magnetic fields could be dynamically important in real jets and lobes since they can contribute some fraction of pressure to the lobes (e.g. \citealt{Konar2009}), and they can affect the stability of the jets (\citealt{Nakamura2001}, \citealt{Tchekhovskoy2016}). The inclusion of cosmic rays (CRs) may also be important, especially if jet-inflated lobes are dominated by CRs. This is because such lobes may not easily exchange energy with the ambient medium, depending on the properties of CR transport (see e.g. \citealt{Ruszkowski}). However, CR physics is still not fully understood, nor is it clear whether CRs are dynamically dominant in real jet-inflated lobes (although they are likely dynamically significant, see e.g. \citealt{Beckmann2022}). Even in the case that magnetic fields and CRs are important for the evolution of jets and lobes, our kinetic jet feedback without magnetic fields and CRs may still quench cooling flows in a manner similar as observed. This is because a large fraction (of order $50\%$ or more) of the energy launched into the jets is quickly transferred to the ICM through bow shocks (e.g. \citealt{Bourne2017}, \citealt{Weinberger2017b}, \citealt{Husko2022a}). These bow shocks are launched by the lobes displacing the ICM and this process should be insensitive to the makeup of the lobes.


\subsection{Dark matter, stars and gas}

The initial conditions for our set-up are discussed in detail in \cite{Nobels2022}. Here we present a summary of the main features of the set-up. The dark matter component is represented with an external \cite{NFW} potential, and its concentration parameter depends on the mass of the system (see \S~\ref{sec:sims}). We include a stellar component in the form of a spherically symmetric Hernquist profile (\citealt{Hernquist1990}). The velocity dispersion of the stellar halo is determined from the Jeans equation (\citealt{Jeans1915}), with the choice of no net rotation.

The main component in our simulations, other than the SMBH, is the gaseous halo, which represents the circumgalactic/intracluster medium (CGM/ICM hereafter). The sound speed of this gas, $c_\mathrm{s}$, is set equal to the circular velocity, $v_\mathrm{c}$, which determines the temperature profile of the halo. Along with the equation of hydrostatic equilibrium, this condition sets the shapes of the pressure and density profiles. We assume that the gas is ideal, with an adiabatic index $\gamma=5/3$. The normalisation of the density profile is determined from the total gas fraction within the $R_\mathrm{500}$ radius, which is calibrated using the results from the BAHAMAS simulations (\citealt{McCarthy2017}), and which reproduce the observed gas fractions. 

In the central regions of the gas halo, the temperature profile is modified such that it can represent a typical profile found in cool-core clusters. This modification is controlled by a free parameter: the minimal central temperature of the gas, $T_\mathrm{0}$. The gas is given a constant fraction of the (radially varying) circular velocity $v_\mathrm{c}$ in the positive $z-$axis direction, such that the dimensionless spin parameter of the halo, $\lambda_\mr{g}=J_\mr{g}/(\sqrt{2}M_\mr{g}V_\mr{200}R_\mr{200}$) (\citealt{Bullock2000}), is equal to the mean value $\lambda=0.05$ for dark matter haloes found in cosmological simulations. Here, $R_\mr{200}$ and $V_\mr{200}$ are the virial radius and the circular velocity at the virial radius of the dark matter halo, respectively, and $M_\mathrm{g}$ and $J_\mathrm{g}$ the total mass and angular momentum, respectively, of the gaseous halo within $R_\mr{200}$. The metallicity of the gas is set to $0.3Z_\odot$ (with $Z_\odot=0.0134$). In the central regions of our gaseous halo, within a radius of $R_\mathrm{res}$, we use a gas particle mass resolution of $m_\mathrm{gas,0}$. The same mass is used to represent the stellar Hernquist component. Beyond $R_\mathrm{res}$ the mass resolution of the gas increases as $m_\mathrm{g}=m_\mathrm{gas,0}(r/R_\mathrm{res})^2$. Using a variable resolution allows for, effectively, higher-resolution simulations to be run. In order to properly resolve the cooling flow and jet feedback, we use a large value of $R_\mathrm{res}=500$ kpc.

\subsection{Simulations}
\label{sec:sims}

We focus on three different systems: the $10^{13}$ $\mr{M}_\odot$, $10^{14}$ $\mr{M}_\odot$ and $10^{15}$ $\mr{M}_\odot$ haloes, where the halo masses are defined as the masses within the virial radius $R_\mathrm{200}$, the radius within which the mean density is 200 times larger than the critical density (assuming $z=0$). The virial radii of the three haloes are $442.7$ kpc, $953.8$ kpc, $2054.8$ kpc, respectively. In terms of virial overdensities computed using mean densities that are 500 times the critical density, the halo masses, $M_{500}$, are $7.79\times10^{12}$ $\mathrm{M}_\odot$, $7.52\times10^{13}$ $\mathrm{M}_\odot$ and $7.16\times10^{14}$ $\mathrm{M}_\odot$, while the virial radii, $R_\mathrm{500}$, are $305.8$ kpc, $651.2$ kpc and $1358.8$ kpc. The concentration parameters of these haloes are $7.2$, $5.6$ and $4.0$. The stellar masses of the galaxies placed in their centres are $10^{11}$ $\mr{M}_\odot$, $2.5\times10^{11}$ $\mr{M}_\odot$ and $6\times10^{11}$ $\mr{M}_\odot$, and the black hole masses are $2.5\times10^{8}$ $\mr{M}_\odot$, $10^{9}$ $\mr{M}_\odot$ and $5\times10^{9}$ $\mr{M}_\odot$, respectively. These systems represent galaxy groups and clusters. The simulations are run for 8 Gyr in the $10^{13}$ $\mr{M}_\odot$ and $10^{14}$ $\mr{M}_\odot$ cases, while the largest system is run for $16$ Gyr due to its longer cooling times. The parameter values used for the simulations presented in this paper are summarised in Table \ref{tab:tab0}.\footnote{The initial conditions can be found online; see the SWIFT repository.}

For each of our three halo masses, we perform a few parameter variations. We simulate each halo at three different mass resolutions, differing by factors of 8 (corresponding to changes in the gravitational softening length, $\epsilon_\mr{g}$, by factors of 2). In the two lower halo mass cases, our highest-resolution simulations have a central particle mass resolution of $m_\mr{gas,0}=10^5$ $\mr{M}_\odot$ and gravitational softening length $\epsilon_\mr{g}=300$ pc. The highest-resolution simulation for our most massive halo is 8 times worse in terms of mass resolution, since it is computationally more expensive. The typical smoothing length of the SMBH in our simulations, as well as the highest-density gas, is $2-3$ times lower than the softening length during strong cooling flows, and around $10$ times higher than that outside the cooling flows. In our highest-resolution simulation, this corresponds to 100 pc and 1 kpc, respectively. These differences arise due to the presence of cold gas or lack thereof.

We vary the initial SMBH spin for each of the three halo masses. Our fiducial spins (directed along the $z-$axis) are $0.2$ in the two lower-mass systems and $0.4$ for the most massive galaxy cluster, corresponding to jet efficiencies of $\approx3\%$ and $\approx12\%$, respectively. Even though we use relatively low values of spin, the jet efficiencies are larger than typically assumed in similar simulations (of order $10^{-3}$, e.g. \citealt{Gaspari2012}, \citealt{Yang2016}, \citealt{Martizzi2019}). This is a result of our assumption that the accretion efficiency is $100\%$, i.e. that there are no disc winds and that all of the matter accreting from the Bondi radius reaches and accretes onto the SMBH. Note that the initial SMBH spin does not only change the efficiency; lower values of spin make the SMBH more susceptible to perturbations in the angular momentum of accreting gas, so the SMBH spin vector will precess more or become reoriented more rapidly. 

For each halo we vary the central temperature of the initial gas distribution, $T_0$. This parameter controls whether the halo being simulated starts off as an analogue of a cool-core cluster (low $T_0$, e.g. $4$ times lower than the virial temperature of the halo), a non-cool-core cluster (high $T_0$, near the virial temperature of the halo), or something in between. For this reason, the choice of the initial central temperature can have a very strong impact on the evolution of the system, as shown by \cite{Nobels2022}.

For the highest-mass halo, we have also performed variations of many other parameters. This includes jet-related parameters such as the launching velocity. We also test cases where the axis along which the jets are launched is fixed to be the $z-$axis; in this situation the jet efficiency is also fixed in time, and we vary this constant efficiency. We found that varying the half-opening angle of the jet does not affect our results (see Appendix \ref{sec:app3}).

We also varied parameters related to the ICM. This includes its total angular momentum, the inclination of the ICM angular momentum vector relative to the initial spin vector of the SMBH, and the metallicity distribution of the ICM. We varied several other unrelated parameters. We found that these variations did not affect the feedback cycle significantly, at least in our most massive halo. We discuss these variations in more detail in Appendix \ref{sec:app3}.


Finally, we varied the scheme with which particles are kicked from the SMBH smoothing kernel. Our standard choice, where we kick particles in the SMBH smoothing kernel that are farthest from it, is compared with that where we kick the closest particles, the ones closest to the spin axis in terms of angular distance, and the ones of lowest density.


\captionsetup[table]{skip=0pt} 
\begin{table*}
\begin{center}
\caption{List of all simulations. In the first three rows we specify the parameters of our fiducial simulations for each of the three halo masses we have simulated. We then specify the ranges of variations of all other parameters in the next three rows. The parameters are, in order: 1) $M_\mr{200}$ - halo mass, 2) $m_\mr{gas,0}$ - central gas resolution in terms of particle mass, 3) $T_\mr{0}$ - central gas temperature, 4) $a_0$ - magnitude of initial SMBH spin, 5) $\epsilon_\mr{j}$ - jet efficiency; constant value or $\epsilon_\mr{j}(a)$, the spin-dependent efficiency given by equation (\ref{eq:epsilon_jet}), 6) $v_\mr{j}$ - jet launching velocity, 7) Scheme - which particles within the SMBH smoothing kernel are kicked from the SMBH, F: farthest, C: closest, S: closest to the axis of the spin vector (in terms of angular distance), L: lowest density. }
\label{tab:tab0}
\end{center}

\begin{tabular*}{1.01\textwidth}{@{\extracolsep{\fill}}lccccccccccccc}
  \hline \hline
  $M_\mr{200}$ $[\mr{M}_\odot]$ & $m_\mr{gas,0}$ $[\mr{M}_\odot]$ & $T_\mr{0}$ [K] & $a_0$ & $\epsilon_\mr{j}$ & $v_\mr{j}$ [$\mathrm{km}\hspace{0.3mm}\mathrm{s}^{-1}$] & Scheme  \\
  \hline \hline
  $10^{13}$ & $10^5$ & $10^{5.75}$ & $0.2$ & $\epsilon_\mr{j}(a)$ & $5\times10^3$ & F  \\
  $10^{14}$ & $8\times10^5$ & $10^{6.75}$ & $0.2$ & $\epsilon_\mr{j}(a)$ & $10^4$ & F \\
  $10^{15}$ & $6.4\times10^6$ & $10^{7.75}$ & $0.4$ & $\epsilon_\mr{j}(a)$ & $3\times10^4$ & F \\
  \hline
  $10^{13}$ & $10^5-6.4\times10^6$ & $10^{5.25}-10^{6.25}$ & $0.1-0.4$ & $\epsilon_\mr{j}(a)$ & $5\times10^3$ & F \\
  $10^{14}$ & $10^5-6.4\times10^6$ & $10^{6.25}-10^{7}$ & $0.1-0.4$ & $\epsilon_\mr{j}(a)$ & $10^4$ & F  \\
  $10^{15}$ & $8\times10^5-5.12\times10^7$ & $10^{7.25}-10^{8}$ & $0.2-0.8$ & $\epsilon_\mr{j}(a), 0.01-1$ & $1.5-6\times10^4$ & F, C, L, S \\
  \hline \hline
\end{tabular*}

\end{table*}

\section{Results: The quenching of galaxies across the mass scale}
\label{sec:sec4}

\subsection{Galaxy group}

In our lowest-mass system, representing an idealised galaxy group with a halo mass of $M_{200}=10^{13}$ $\mathrm{M}_\odot$, we find that regardless of the initial SMBH spin, mass resolution or initial central gas temperature, the evolution of the system is similar. Initial cooling due to the presence of a cool core leads to a strong jet episode, which subsequently turns off any significant cooling during the next $8$ Gyr of evolution. Our results are similar to those found by \cite{Nobels2022} for the same initial conditions using thermal AGN feedback instead of jets. 
We find that the Eddington-normalised accretion rate $\dot{m}$ reaches peak values of $\approx0.01$ during the initial cooling flow, but only for the low initial spin case ($a_0=0.1$) and low initial central gas temperature case ($T_0=10^{5.25}$ K). This lasts only for several Myr, after which the accretion rate falls well below that value. In other cases, the accretion rate is always well below $0.01$, indicating that the jet-efficient, thick disc regime is applicable in these simulations.

\begin{figure}
\includegraphics[width=1.01\columnwidth, trim ={0 0 0 0},clip]{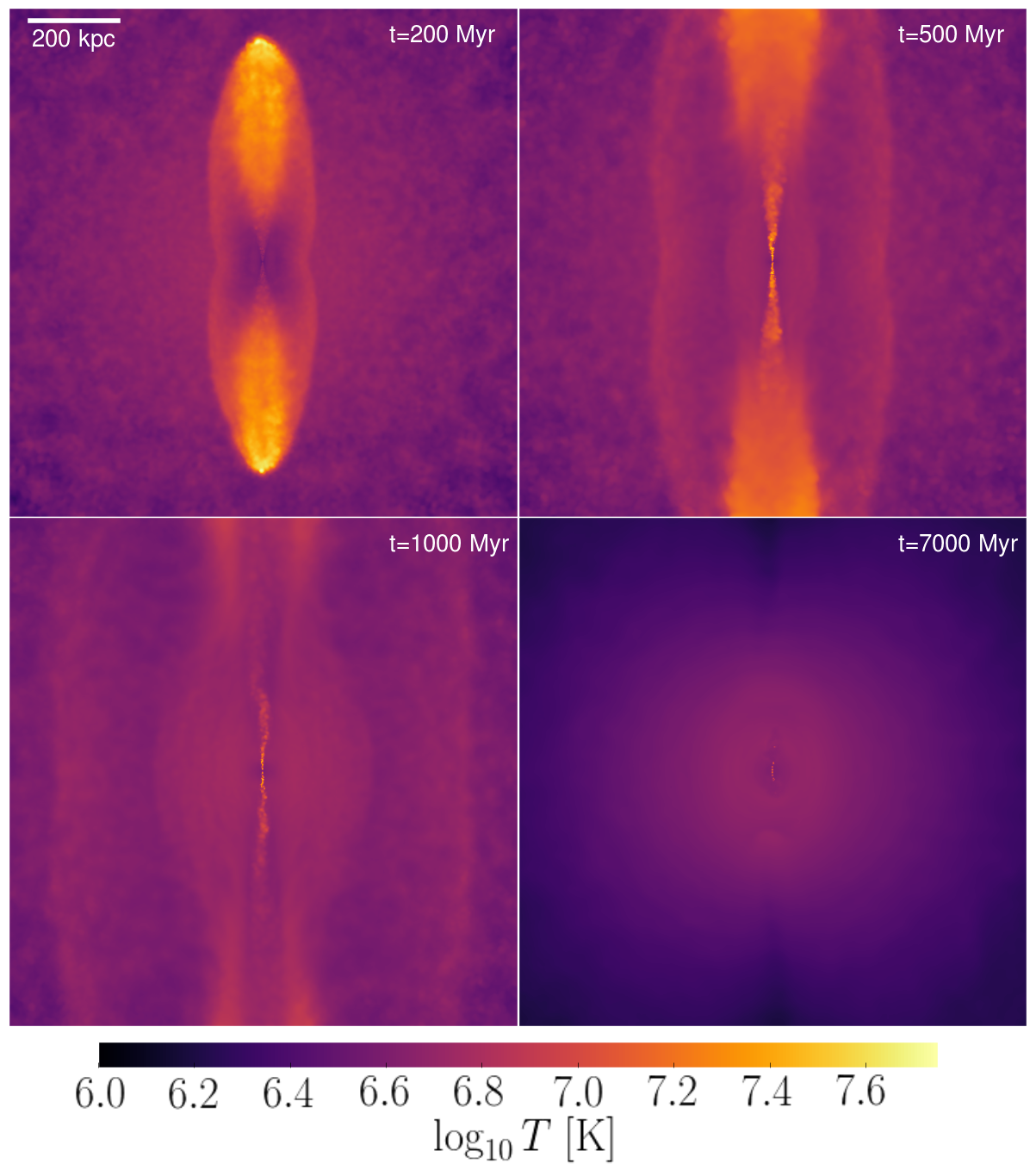}
\caption{Gas temperature projections (mass-weighted mean) in our fiducial simulation (see Table \ref{tab:tab0} for details) of the galaxy group halo ($M_\mr{200}=10^{13}$ $\mr{M}_\odot$) at different times. The images show a region 1 Mpc across and 200 kpc in depth. They show initially strong jets that inflate lobes and launch bow shocks, a weaker, precessing jets at intermediate times, and a very weak jet at late times.}
\label{fig:fig1}
\end{figure}

Fig. \ref{fig:fig1} shows the temperature of the gas in our highest-resolution simulation with our fiducial jet launching parameters. We see ellipsoidal lobes being inflated in the first two snapshots. In the third snapshot we see a weak jet, resembling an FRI source (\citealt{Fanaroff}). This jet is weakly precessing due to the chaotic nature of the angular momentum of accreting gas and since redirecting the spin vector by a few degrees requires very little accretion.  The spin value of the SMBH stays very similar to the initial one. The last snapshot shows the system at late times. By this point, the jet power has reduced even more, but it is still non-zero. The system is kept in a steady state by these very weak jets.

\begin{figure*}
\includegraphics[width=1.01\textwidth, trim = 0 10 0 0]{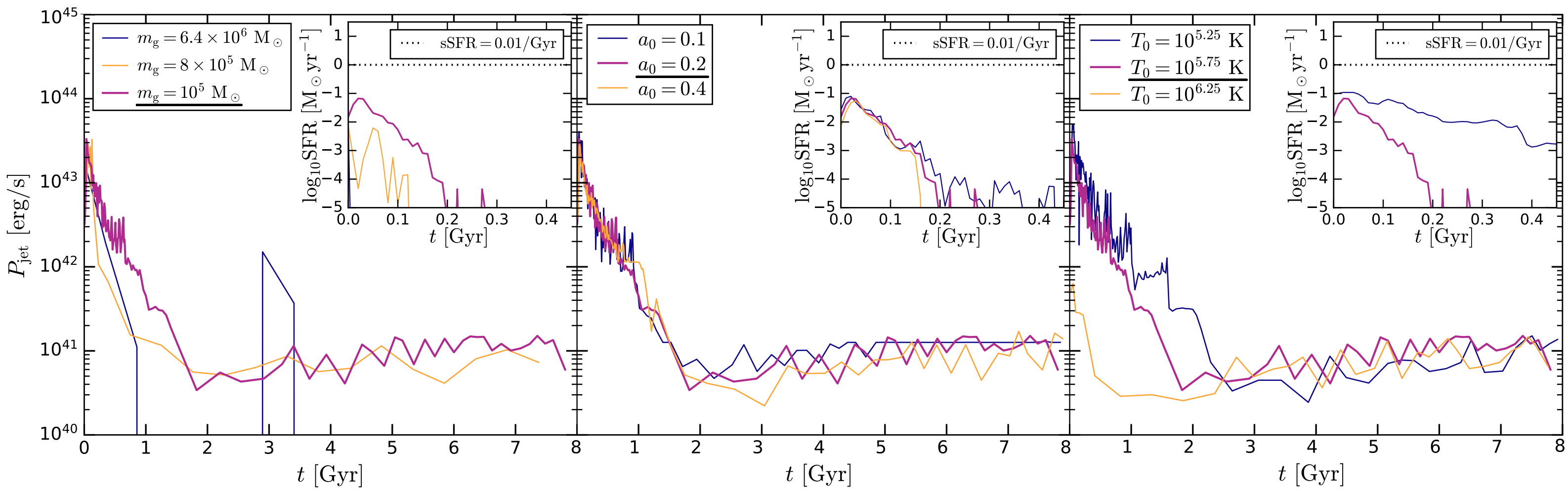}
\caption{Jet power and star formation rate in the $M_\mr{200}=10^{13}$ $\mr{M}_\odot$ simulations with varying resolution (left), initial SMBH spin (middle) and initial central temperatures (right). The details of the fiducial case, relative to which these variations are made, are given in Table \ref{tab:tab0} (purple line in each panel, underlined parameter in each panel legend). The dotted black lines represent the upper limit of the specific star formation rate required to classify a galaxy as quenched.}
\label{fig:fig2}
\end{figure*}

In order to quantify jet feedback, we focus on the time dependence of jet powers and star formation rates (cold gas masses follow the SFR very closely). Fig. \ref{fig:fig2} shows the jet powers and star formation rates in our $M_\mr{200}=10^{13}$ $\mr{M}_\odot$ simulations, with varying numerical resolution, initial SMBH spin and initial central temperature. These are all very similar, in that there is an initial jet and SFR episode, with the SFR being fully quenched by $t=0.5$ Gyr. The jet power gradually reduces after reaching a peak within the first $0.5$ Gyr. The powering down of the jets is completed by $2$ Gyr in all cases. For the remainder of the simulations, the jet powers remain close to their average values, indicating that the system has reached a quenched steady state. 

From the left-hand panel, we see that jet powers converge onto the same time dependence across different resolution levels. During the initial jet episode, the three simulations have a very similar jet power. The highest-resolution simulation features a more protracted decrease from the peak, possibly because the jets in that simulation can travel to farther distances and thus heat local gas less effectively. The highest-resolution simulation is also the most variable, as expected due to the finer sampling of energy injection. Star formation is present only during the peak of the initial episode and only in the two higher-resolution runs, and it increases with resolution. After the initial jet episode, the lowest resolution run is so noisy that it features only a few jet kicking events around $t=3$ Gyr. The two higher-resolution runs appear converged onto a fairly constant jet power after $2$ Gyr, with a value of $P_\mr{jet}\approx10^{41}$ $\mathrm{erg}\hspace{0.3mm}\mathrm{s}^{-1}$. The lower-resolution simulation is less variable in this period due to courser sampling.

From the middle panel, we see that the details of the quenching are very similar regardless of initial SMBH spin (which is, in this case, a proxy for jet efficiency, since spin varies very little during the simulations, and jet efficiency varies as $\epsilon_\mr{j}\propto a^2$ at small spin values, see \S~\ref{sec:jet_eff}). The main difference is that the highest spin case appears less variable during the initial jet episode. This is likely due to the SMBH being able to react more quickly to gas cooling, by launching a pair of particles earlier and thus preventing buildup of too much cold gas.

From the right-hand panel we see that haloes with lower initial central temperatures show more energy injection from jets, as well as more star formation, as expected due to higher rates of gas cooling. Star formation is quenched successfully in all three cases. By $t=3$ Gyr, all simulations converge onto the same jet power as in the previous cases, including the one where the central temperature is close to the virial temperature. This likely indicates that all of the gas that differs between the initial profiles is effectively heated or ejected from the central regions of the halo. Since the spin remains constant ($a=0.2$), a constant jet power implies that the accretion rate is the same between the different simulations. This accretion rate corresponds to Bondi growth directly from the hot halo. These simulations indicate that 'hot accretion' is sufficient to keep the galaxies quenched, at least in systems with $M_{200}=10^{13}$ $\mr{M}_\odot$.

\subsection{Low-mass galaxy cluster}

In the low-mass galaxy cluster case, with a halo mass of $M_{200}=10^{14}$ $\mathrm{M}_\odot$ halo, we find that hot halo accretion is not sufficient to keep the central galaxy quenched after the first jet episode. Instead, the galaxy experiences multiple episodes of jet activity and star formation; in each episode, the jets are fed by cold gas. We find that the accretion rate $\dot{m}$ never exceeds values of $0.03$, indicating that our jet feedback model is applicable in these simulations.

\begin{figure}
\includegraphics[width=1.01\columnwidth, trim ={0 0 0 0},clip]{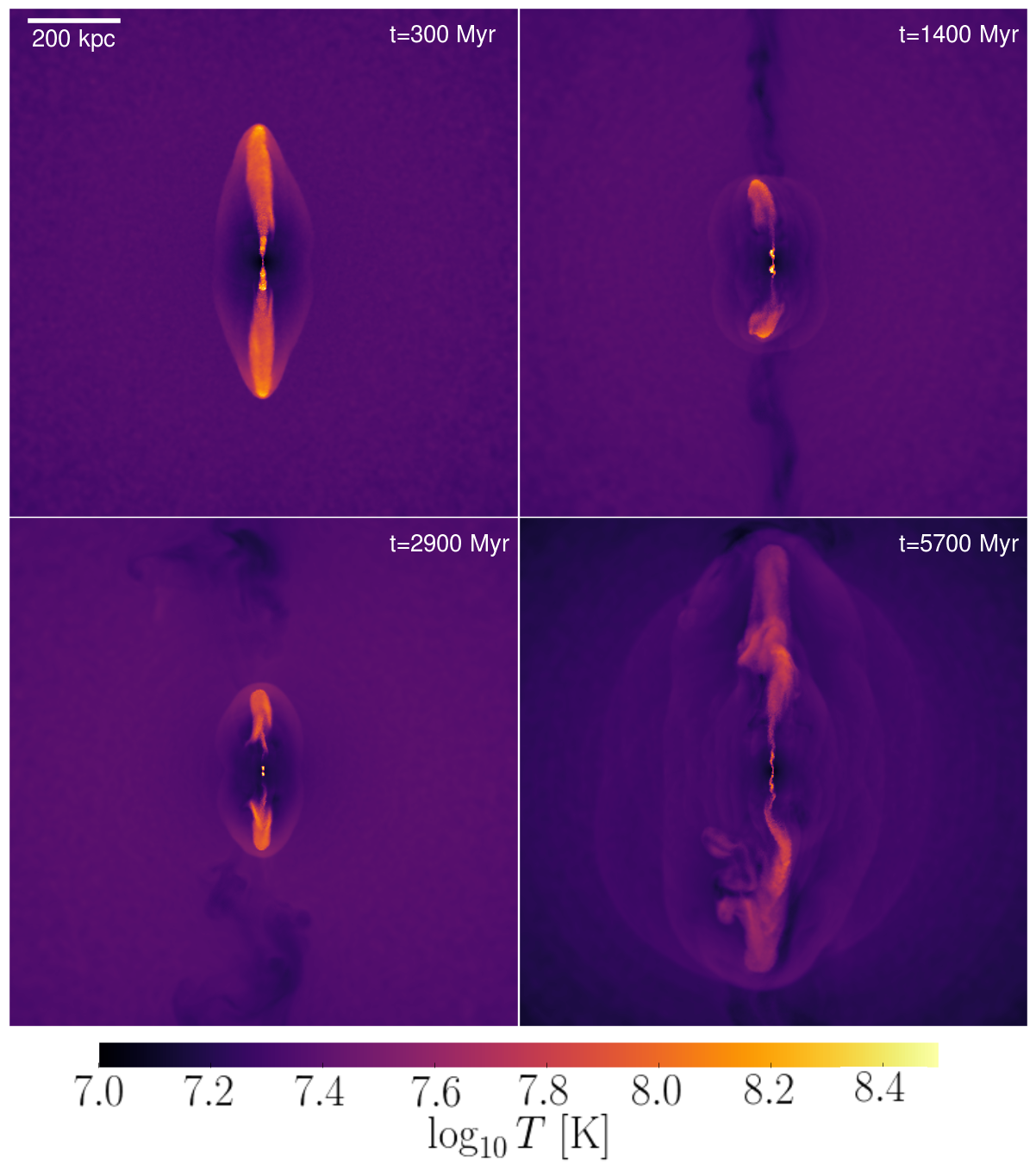}
\caption{Gas temperature projections (mass-weighted mean) in our fiducial simulation (see Table \ref{tab:tab0} for details) of the low-mass galaxy cluster ($M_\mr{200}=10^{14}$ $\mr{M}_\odot$) at different times. The images show a region 1 Mpc across and 200 kpc in depth. They show the variety of jet morphologies featured during and between multiple cooling flows. The video version of this plot is available to view at \href{https://youtu.be/Edf2hS7HU70}{https://youtu.be/Edf2hS7HU70}.}
\label{fig:fig3}
\end{figure}%

Fig. \ref{fig:fig3} shows visualisations of jets in the low-mass galaxy cluster at various times, in our highest-resolution simulation ($m_\mr{g}=10^5$ $\mr{M}_\odot$), with our fiducial jet parameters. Since we use the same gas mass resolution as for the galaxy group simulations, and the typical jet powers are significantly larger, the jets appear better resolved. These snapshots highlight the different jet morphologies seen throughout this simulation. In the first snapshot we show the initial jet episode. We see two ellipsoidal lobes, as well as bow shocks propagating through the halo. The hottest gas is near the jet head, as well as near the jet base. This indicates that the jets have features akin to both FRI and FRII jets (\citealt{Fanaroff}). 


In the second snapshot we show the aftermath of a second, weaker episode that occurs $\approx1$ Gyr after the first one. The third snapshot shows an episode of a similar power after a third episode. In both the second and third snapshot, there are signs of low-temperature gas ahead of the bubbles inflated by these weaker jet episodes. This low-temperature gas is a result of uplift of low-entropy gas caused by the first, strong jet episode. We discuss this gas uplift in \S~\ref{sec:uplift}. The last snapshot shows the complex morphology caused by a precessing jet that is also varying in power during its episode, causing multiple distinct bow shocks. These jets and bow shocks are also interacting with the infalling low-entropy gas that was previously uplifted, further complicating the picture. 

\begin{figure*}
\includegraphics[width=1.01\textwidth, trim = 0 10 0 0]{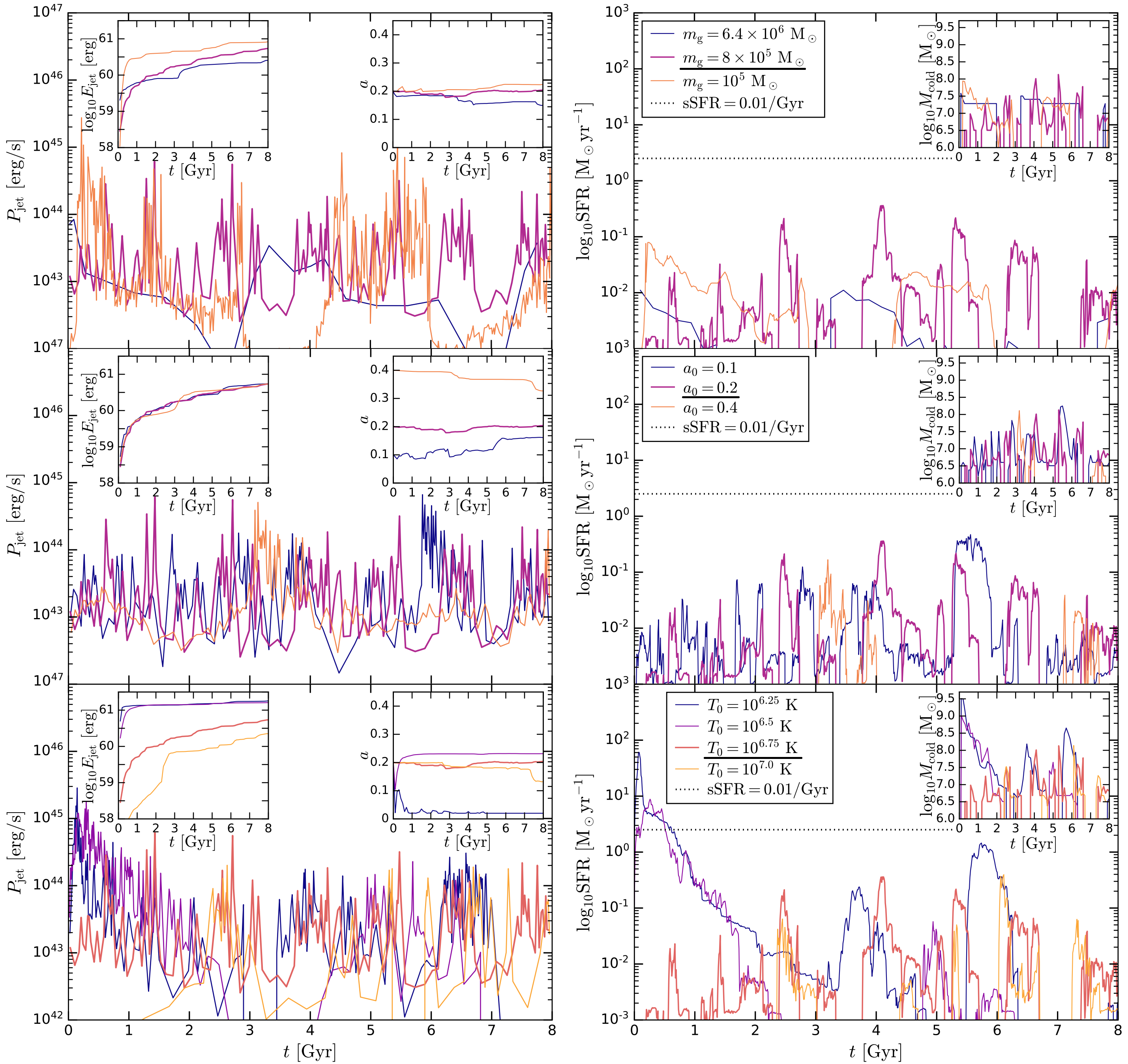}
\caption{Time dependence of the quenching/feedback process in the low-mass galaxy cluster simulations ($M_\mr{200}=10^{14}$ $\mr{M}_\odot$) with varying mass resolution (top row), initial SMBH spin (middle row) and initial central temperatures (bottom row). The left-hand panels show the jet power, while the right-hand panels show the star formation rate. Insets in the left-hand panels show the injected jet energy and magnitude of SMBH spin. The insets in the right-hand panels show the cold gas mass. The details of the fiducial case, relative to which these variations are made, are given in Table \ref{tab:tab0} (purple and orange lines in the top/middle and bottom panels, respectively; underlined parameter in each panel legend). The dotted black lines represent the upper limit of the specific star formation rate required to classify a galaxy as quenched.}
\label{fig:fig4}
\end{figure*}%

Fig. \ref{fig:fig4} shows the time dependence of jet powers and star formation rates in our low-mass galaxy cluster simulations with varying parameters (the same ones as for the galaxy group case, shown in Fig. \ref{fig:fig2}). We also include plots of total injected energy, cold gas mass (cold gas here meaning gas with $T<2\times10^4$ K) and magnitude of SMBH spin. In all cases, jet feedback leads to successful quenching.

In the top panels we show the results of varying the numerical resolution. Overall, increasing the resolution leads to more energy injection from jets, as well as more variability in the jet power. The amounts of cold gas do not increase with resolution, while the SFR increases only from the lowest to the intermediate-mass case. In the two lower-resolution cases, the peaks in the SFRs are well correlated with peaks in the jet power, indicating that the cool gas feeding the jets is also star-forming. In the highest-resolution case, the cold gas mass and SFR is less variable, possibly due to the jets affecting the cold gas to a lesser degree at higher resolutions. The cold gas masses that we find, of order $10^7-10^8$ $\mr{M}_\odot$, are consistent with observations of massive elliptical galaxies (\citealt{Georgakakis2001}, \citealt{OSullivan2015}).

In the middle panels we see how the feedback depends on the initial SMBH spin. Unlike in the low-mass simulations, the spin changes somewhat. The medium-spin simulation shows no spin evolution, while in the low-spin simulation, the SMBH is spun up from $a=0.1$ to $a=0.15$. The higher-spin simulation features spindown, from $a=0.4$ to $a=0.32$ by $t=8$ Gyr. This spindown is a result of jet activity; in the thick disc regime, jet launching causes a decrease in spin for SMBHs with spins above $\approx0.25$ (see the bottom panel of Fig. \ref{fig:fig0}). Although we do not show these results here, we find that the direction of the spin vectors is very well aligned with the $z-$axis, with only the low-spin case showing a small deviation ($10\degree$) from the initial direction.

The simulations with $a_0=0.1$ and $a_0=0.2$ both feature fairly variable jet powers. The peaks of jet activity are very well correlated with peaks in the SFR and cold gas mass. The galaxies are quenched at all times, with cold gas masses reaching values of up to $10^8$ $\mr{M}_\odot$. Outside the strongest SFR/jet episodes, we find small cold gas reservoirs with $M_\mr{cold}<10^7$ $\mr{M}_\odot$ throughout most of the simulation. The jet powers are always above $10^{42}$ $\mathrm{erg}\hspace{0.3mm}\mathrm{s}^{-1}$, which represents the minimum jet power from hot halo accretion. The case with $a_0=0.4$ is less variable than the other two. The jet power exhibits two peaks (at $t=3$ Gyr, and at the very end of the simulation), which coincide with periods when cold gas mass is present, and when stars are being formed. The cold gas masses and SFRs are lower than in the cases with lower initial SMBH spin. This is probably a result of the jet being able to react to an accumulation of cold gas more rapidly (due to higher jet efficiencies), thus promptly shutting off a cooling flow. At the same time, the hot halo accretion launches stronger jets (i.e. the minimum in the jet power is higher, nearer to $10^{43}$ $\mathrm{erg}\hspace{0.3mm}\mathrm{s}^{-1}$, rather than $10^{42}$ $\mathrm{erg}\hspace{0.3mm}\mathrm{s}^{-1}$), which results in less cooling during eventual cooling episodes. During the very beginning of the simulation, this is likely what prevented any gas from cooling quickly and launching an initial jet episode. Despite the qualitative differences discussed so far, the total injected energy is very similar in all three cases. 

In the bottom panels we show results of varying the initial central temperature. A case with higher initial temperature than fiducial, close to the virial temperature, takes a longer time to show any jet/star formation activity, but even in this case there are jet/star formation cycles. As expected, lower initial central temperatures lead to more cold gas (exceeding $10^9$ $\mr{M}_\odot$), more star formation and stronger jet activity. In the two lowest-temperature cases, the simulations feature strong initial jet episodes, similar to the lower mass halo. The jet powers peak at $10^{45}$ $\mathrm{erg}\hspace{0.3mm}\mathrm{s}^{-1}$, and the SFRs reach $100$ $\mr{M}_\odot\mr{yr}^{-1}$ in the lowest-temperature case. Unlike the galaxy group case, both of these simulations also feature further jet and star formation episodes later on. However, in the later episodes the jet powers are weaker, and the SFRs are low enough to consider the galaxies quenched. Both simulations feature significant spin evolution. There is significant initial SMBH spindown, with the lowest-temperature case settling down to a very low spin of $a=0.03$ and misaligned relative to the $z-$axis (not shown here). The somewhat higher temperature case features spinup back to around $a=0.2$ during the initial jet episode, and the angle between the spin vector and $z-$axis is small throughout the simulation. This is expected since large amounts of cooling generally result in a cold circumnuclear disc. In the lowest-temperature case, it is possible that this did not occur, and the SMBH was spun down into an effectively random direction, because there was sufficient cooling in the very centre of the gaseous halo (where the angular momentum of the gas is lower).

Our results for the low-mass galaxy cluster are overall similar to \cite{Nobels2022} for the same system using thermal feedback. However, we find that the jets quench cooling more effectively, leading to less star formation. The quenching is also less protracted. Furthermore, the jets are able to quench haloes with lower initial central gas temperatures. Finally, we find that our cooling and jet episodes are largely non-periodic, while \cite{Nobels2022} find periodicity. This difference is likely a result of varying efficiency in the jet case. 

\cite{Beckmann2019} performed simulations similar to ours (including AGN jet feedback and SMBH spin evolution) and focused on a set-up of an idealised Perseus-like galaxy cluster (which they assumed to have a halo mass of $3.4\times10^14$ $\mathrm{M}_\odot$, this may be more comparable to our high-mass galaxy cluster simulations$-$see next subsection). They found much higher cold gas masses and star formation rates than we do ($M_\mathrm{cold}=10^{10}-10^{11}$ and $\mathrm{SFR}=0-1000$ $\mathrm{M}_\odot\mathrm{yr}^{-1}$, respectively). Such a lack of strong quenching is not necessarily surprising since the Perseus cluster is a cool-core cluster (\citealt{Schmidt2002}), as well as having a central SMBH which is relatively undermassive for its host halo, by an order of magnitude (\citealt{Sani2018}).

The recurrence time between SFR episodes found by \cite{Beckmann2019} is of order $0.1-0.2$ Gyr instead of $1-2$ Gyr as in our case (even when we compare their simulations with our cool-core simulations). This suggests that their jets heat the ICM more locally instead of traveling to the outer reaches of the halo$-$this interpretation is in line with the distances reached by the jets in the two sets of simulations (tens of kpc in their case versus hundreds of kpc in ours). This means that the closest gas that has not been effectively heated lies at smaller radii in their case than in ours. Such gas has shorter cooling times, leading to a shorter recurrence interval between cooling flows. The difference in the distances reached by the jets may be due to numerical resolution. \cite{Beckmann2019} resolve the ICM with a cell mass of $m_\mathrm{gas}=3.5\times10^{9}$ $\mathrm{M}_\odot$ (more than a factor of $10^3$ poorer resolution than in our case). They resolve their jets to a much better degree (up to $\approx100$ pc, which is in turn much better than our resolution in the jets, of order 1 kpc), but this is progressively derefined as the gas launched by the jets ages (with an exponential decay time of $10$ Myr). On the order of several tens of Myr, the jets probably quickly deposit their energy into the local ICM as they become poorly resolved, thus only being able to reach distances of tens of kpc.

\subsection{High-mass galaxy cluster}

We now turn to our most massive test case, an idealised high-mass galaxy cluster with a halo mass of $M_{200}=10^{15}$ $\mathrm{M}_\odot$. We find multiple episodes of gas cooling and jet activity in this system, similar to the low-mass galaxy cluster. However, in this case, even with our fiducial initial central temperature, the cooling flows are strong enough to induce significant SMBH growth, and therefore also changes in SMBH spin (both its magnitude and direction). 

We find that the accretion rate $\dot{m}$ occasionally exceeds $0.03$, i.e. at those times our thick disc and jet model is unrealistic. Instead, the SMBH should enter the radiatively-efficient thin disc regime (\citealt{ShakuraSunyaev1973}). We find that these periods are relatively short ($<5$ Myr in every case). However, in a realistic simulation where the SMBH switches between the regimes depending on the accretion rate, it is possible that these periods of high accretion rates may be longer. This is because once the SMBH enters the radiatively-efficient regime, it is likely that the thermal feedback associated with radiation is less effective at quenching the cooling flow, which would further exacerbate an ongoing increase in accretion rate. We leave a study of the interplay between thermal and jet feedback in such a scenario for a future study.

\begin{figure}
\includegraphics[width=\columnwidth, trim = 0 10 0 0]{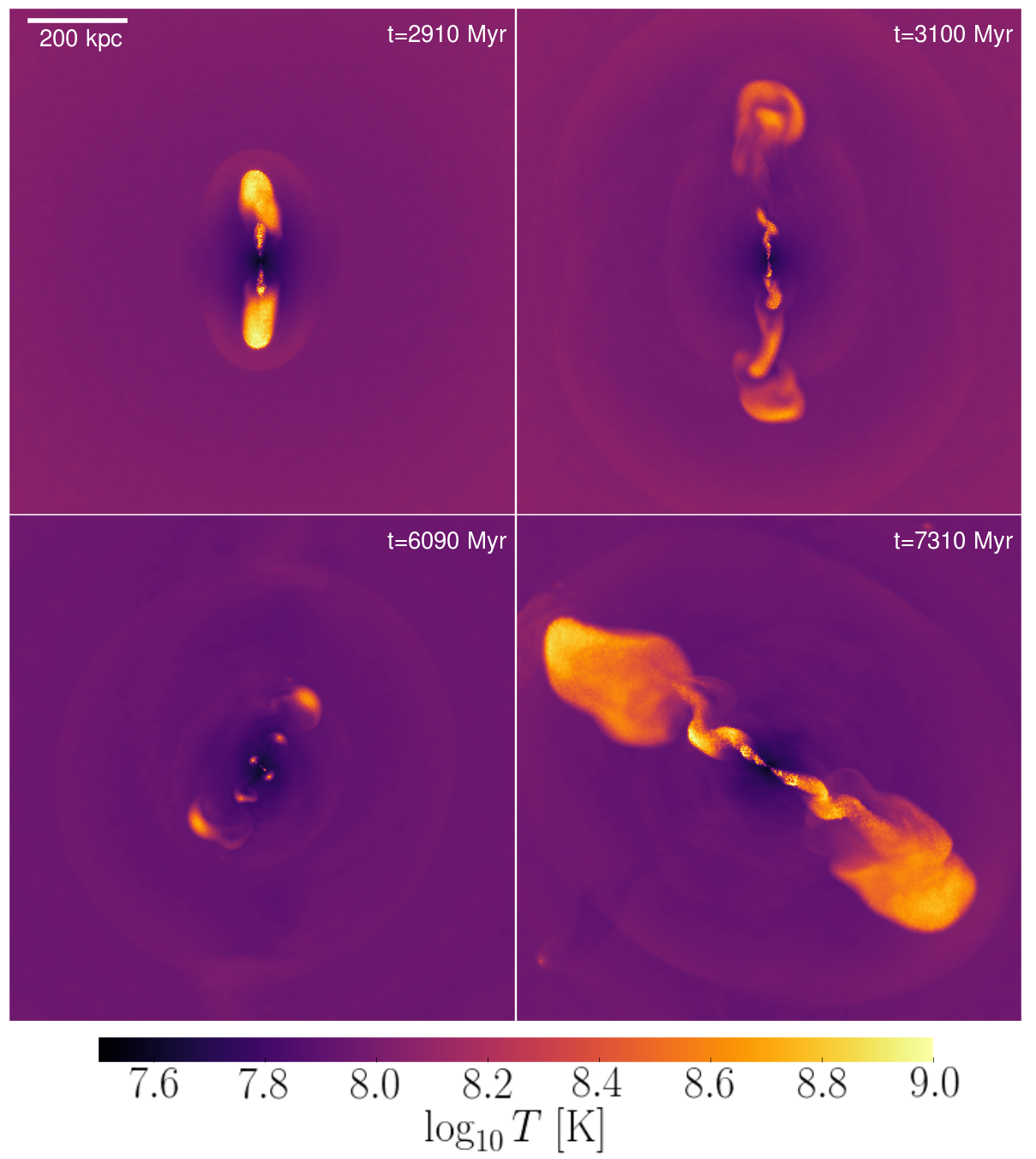}
\caption{Gas temperature projections (mass-weighted mean) in our fiducial simulation (see Table \ref{tab:tab0} for details) of the high-mass galaxy cluster ($M_\mr{200}=10^{15}$ $\mr{M}_\odot$) at different times. The images show a region 1.2 Mpc across and 400 kpc in depth. They show the variety of jet and bubble morphologies in this simulation, as well as jet reorientation. The video version of this plot is available to view at \href{https://youtu.be/2herQHMnrZs}{https://youtu.be/2herQHMnrZs}.}
\label{fig:fig5}
\end{figure}

Fig. \ref{fig:fig5} shows the gas temperature at various times in our highest-resolution simulation of the high-mass galaxy cluster. In the first snapshot we see jets inflating a pair of bubbles, very close to the $z-$axis. In the second snapshot, the spin vector is still aligned with the $z$ direction, and we see a highly precessing active jet, as well as lobes/bubbles from a previous pair of episodes (which are blending into a single one in the top half). In the third snapshot, we see three pairs of bubbles, the outermost two of which are in the same direction, while the innermost pair appears perpendicular to those. None of these are in the $z$ direction, with the spin axis of the SMBH having been changed. The last snapshot shows a strong jet being launched from the feeding off of a circumnuclear disc, which results in a long-lived jet with a well-defined direction, but also showing clear precession. These jets are 700 kpc long (each). 

\begin{figure*}
\includegraphics[width=1.01\textwidth, trim = 0 10 0 0]{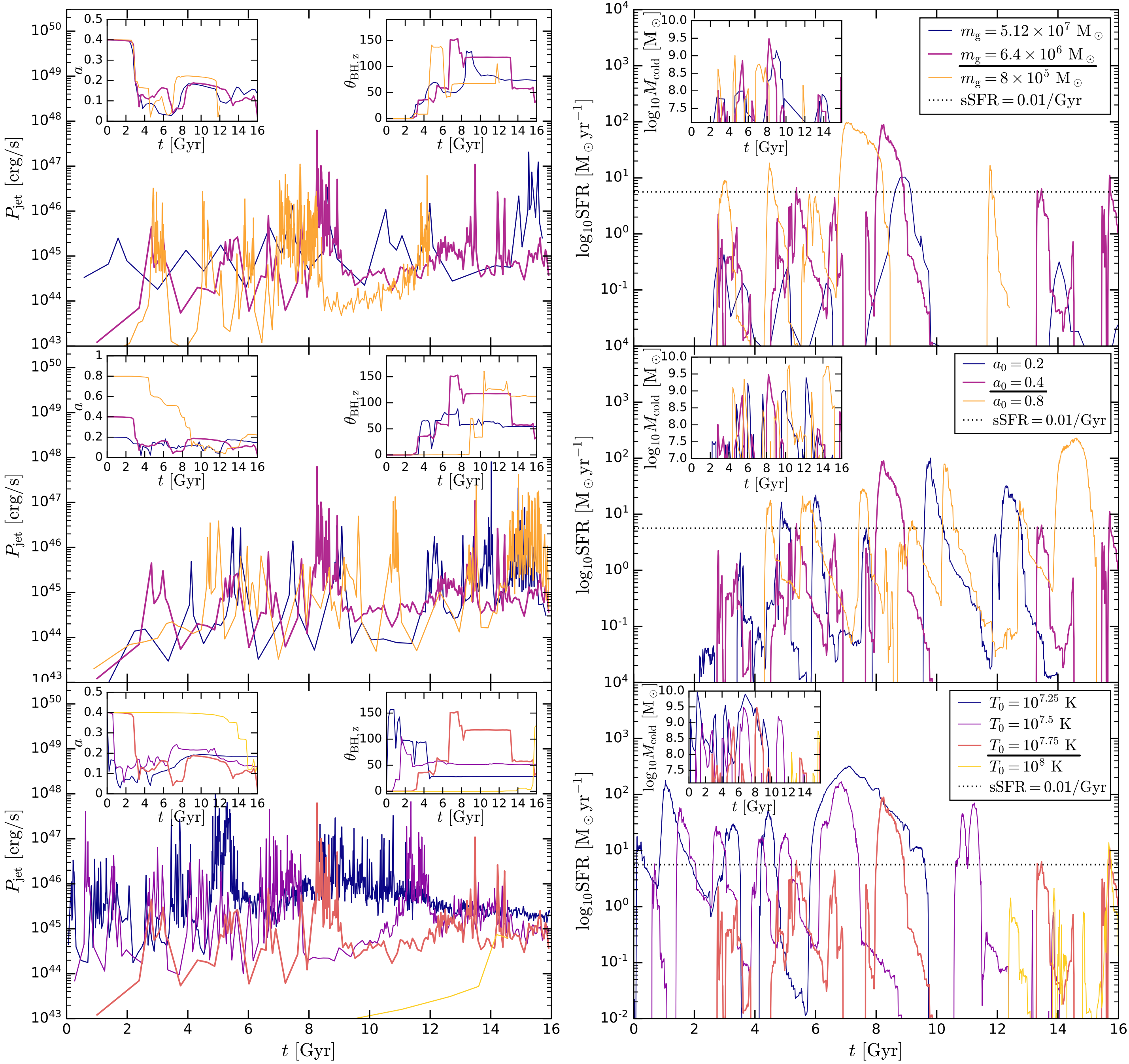}
\caption{Time dependence of the quenching/feedback process in the high-mass galaxy cluster simulations ($M_\mr{200}=10^{15}$ $\mr{M}_\odot$) with varying mass resolution (top row), initial SMBH spin (middle row) and initial central temperatures (bottom row), as per the legends in the right-hand panels. The left-hand panels show the jet power, while the right-hand panels show the star formation rate. Insets in the left-hand panels show the magnitude of SMBH spin and the angle between the spin vector and $z-$axis. The insets in the right-hand panels show the cold gas mass. The details of the fiducial case, relative to which these variations are made, are given in Table \ref{tab:tab0} (purple and orange lines in the top/middle and bottom panels, respectively; underlined parameter in each panel legend). The dotted black lines represent the upper limit of the specific star formation rate required to classify a galaxy as quenched. The yellow line in the top panel is discontinued at $\approx12$ Gyr since this simulation (the highest-resolution one) was not run for the full $16$ Gyr.}
\label{fig:fig6}
\end{figure*}

Fig. \ref{fig:fig6} shows a similar set of plots as Fig. \ref{fig:fig4}, but for this most massive system. We also show plots of the misalignment angle between the SMBH spin vector and the $z-$axis, instead of the total injected energy. We see that the initial cooling flow takes a much longer time to develop (to the point of a non-zero cold gas mass reservoir/star formation), due to longer cooling times of the initial gas profile. In our fiducial case with an initial central temperature of $10^{7.75}$ K, this takes $3-4$ Gyr. Overall, we again find successful quenching of star formation, with multiple cycles of feedback. The peak jet powers approach $10^{47}$ $\mathrm{erg}\hspace{0.3mm}\mathrm{s}^{-1}$, peak cold gas masses approach $10^{10}$ $\mathrm{M}_\odot$, and peak SFRs reach values up to $100$ $\mr{M}_\odot\mr{yr}^{-1}$. The peaks in SFRs are often large enough for the central galaxies to not be considered quenched. However, this is not a 'problem' per se, as many observations of central galaxies in clusters find similarly large cold gas mass reservoirs and SFRs (e.g. \citealt{Odea2008}), sometimes up to $10^{11}$ $\mr{M}_\odot$ and $1000$ $\mr{M}_\odot \mr{yr}^{-1}$, respectively (\citealt{Edge2001}, \citealt{McDonald2015}, \citealt{Castignani}, \citealt{OSullivan2021}). These huge cooling flows are not found in the absence of jet feedback, but are instead correlated with it (\citealt{Hlavacek-Larrondo2012}, \citealt{McNamara2014}, \citealt{Russell2017}).

In the top panels of Fig. \ref{fig:fig6} we show jet powers and SFRs with varying numerical resolution. All three simulations inject similar amounts of energy, but higher-resolution simulations show larger variability and clearly distinct episodes. Smaller jet/cooling episodes differ in timing and peak jet power/SFR, but all three simulations show a similar episode at $t\approx8$ Gyr. The cold gas masses appear well converged with resolution, while the SFR grows by about an order of magnitude when resolution is increased from $m_\mathrm{g}=5.12\times10^7$ $\mathrm{M}_\odot$ to $m_\mathrm{g}=6.4\times10^6$ $\mathrm{M}_\odot$, but is converged down to $m_\mathrm{g}=8\times10^5$ $\mathrm{M}_\odot$. In all three cases the SMBH is spun down during the initial jet episode, and its spin varies throughout the simulation between values of $0$ and $0.25$. The spin vector becomes misaligned in all three cases, but there is no sign of less misalignment at higher resolutions. This indicates that the misalignment is not an effect of poor sampling of the gas distribution, but rather a physical effect. However, in our highest-resolution simulation, the cold gas reaches peak masses of $10^9$ $\mr{M}_\odot$, which means that it is resolved by about $1000$ particles at most. This may not be enough to draw any strong conclusions about the morphology of the cold gas, and therefore about the evolution of the spin vector in terms of direction. It is possible that higher resolutions might result in fewer but longer-lived cooling episodes resulting in cold gaseous discs.

The middle panels of Fig. \ref{fig:fig6} show results of simulations with varying initial SMBH spin. We see that the the SMBH is spun down somewhat during the first cooling/jet cycle in all three cases, and the behaviour of spin is similar after the spindown. In the highest-spin case ($a_0=0.8$), the SMBH gets spun down to $0.6$ during the initial cooling episode at $t=3.5$ Gyr, and then it gets completely spun down in the second cooling episode at $t=8-9$ Gyr. For the remainder of the simulation, in all three cases the spin takes on values between $0$ and $0.2$, with the latter maximal value near the equilibrium spin due to jet spindown. The case with largest initial SMBH spin initially shows a smaller angle between the spin vector and the $z-$axis, since it is harder to steer it off into a different direction. On the other hand, the cases with $a_0=0.2$ and $a_0=0.4$ both have spin vectors that are pointed in a different direction almost immediately during the first jet episode. Despite the differences in spin, all three cases exhibit a similar total injected jet energy as a function of time (not shown here), as well as similar star formation rates.

From the bottom panels we see the effects of varying the initial central temperature. The (relative) changes are similar to the low-mass galaxy cluster case. As expected, decreasing the temperature leads to more energy injection, cold gas and star formation, as well as more rapid spindown and reorientation. The case with the lowest initial central temperature ($T=10^{7.25}$ K) has peak jet powers of a few times $10^{47}$ $\mathrm{erg}\hspace{0.3mm}\mathrm{s}^{-1}$, corresponding to some of the strongest observed jets (\citealt{Kino2005}). The SFR reaches peaks of a few times $100$ $\mr{M}_\odot \mr{yr}^{-1}$, which corresponds to SFRs of central galaxies in clusters with some of the strongest cooling flows, such as the Phoenix cluster (\citealt{McDonald2015}). The galaxy would be considered non-quenched most of the time. However, even in this case, after $11$ Gyr the galaxy is completely quenched. In the case with the somewhat higher initial central temperature of $T_0=10^{7.5}$ K, the SFR is relatively high during the first $8$ Gyr, but the galaxy is again quenched after that, with the exception of another episode at $t=11$ Gyr. In the case where the central temperature is close to the virial temperature, there is almost no cold gas, star formation or jet activity.

\section{Jet feedback in more detail}
\label{sec:sec5}

In the previous section we focused on the general morphology of self-consistent jets, as well as the details of the feedback cycle as measured through the jet power and SFR. Here we will look at some secondary features of these jets and their feedback. We focus on the most massive halo that we have simulated, the high-mass galaxy cluster ($M_\mr{200}=10^{15}$ $\mr{M}_\odot$).

\begin{figure*}
\includegraphics[width=0.99\textwidth, trim = 0 10 0 0]{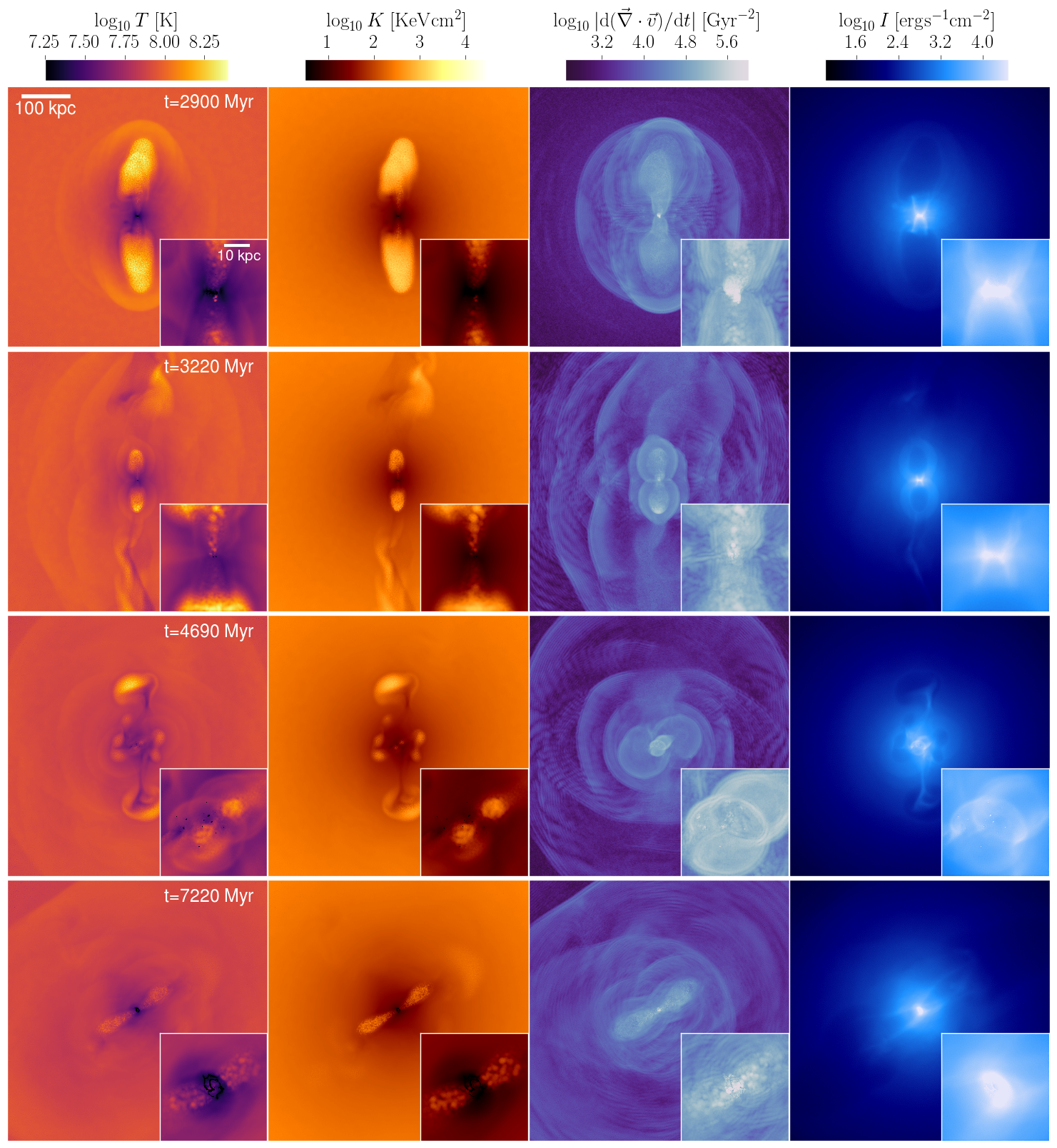}
\caption{Gas properties (mass-weighted means) at different times in our fiducial simulation (see Table \ref{tab:tab0} for details)  of the high-mass galaxy cluster ($M_\mr{200}=10^{15}$ $\mr{M}_\odot$). Each image is 800 kpc across, and shows slices 200 kpc in depth. Insets show a zoom-in of the central 50 kpc (10 kpc in depth). From left to right we show the temperature, entropy, shock/sound wave indicator and X-ray surface brightness (see text for details). From top to bottom we show different times. The video version of the X-ray plots is available to view at \href{https://youtu.be/113F4ndbm1c}{https://youtu.be/113F4ndbm1c}.}
\label{fig:fig8}
\end{figure*}

Fig. \ref{fig:fig8} shows visualizations of gas properties (in slices, and also including zoom-ins, see the caption) in our highest-resolution simulation ($m_\mr{g}=8\times10^{5}$ $\mr{M}_\odot$) of the massive halo, through its temperature, entropy, magnitude of the time derivative of the velocity divergence (this quantity is a shock/sound wave tracer) and the X-ray surface brightness\footnote{We calculate the X-ray surface brightness as appropriate for the ACIS detector on board the Chandra space telescope (\citealt{ACIS}) by using its effective area as a function of photon energy (corresponding to $0.2-7$ keV), which is convolved with a spectrum of bremsstrahlung cooling in an optically thin medium. The presence of metals is accounted for in the total cooling function, but metal lines are not included in the spectrum.}. We have chosen these quantities since they highlight some of the main features of interest. The particular times (snapshots) shown were chosen for a similar reason.

\subsection{Uplift of filaments}
\label{sec:uplift}

The jets in our simulations inflate bubbles that rise buoyantly due to gravity. Such bubbles can be seen clearly in the first three snapshots shown in Fig. \ref{fig:fig8}, in maps of all of the properties to varying degree. We find that the rise of such bubbles is ubiquitously followed by the rise of colder, ambient medium in the form of filaments that connect the base of the bubbles to the centre of the halo. This gas is visible in the second and third snapshots in the temperature maps due to its low temperature ($\approx10^7$ K), and the entropy maps show that the filaments are of a somewhat lower-entropy relative to the rest of the ambient medium. The filaments are also visible in the X-ray surface brightness, which shows actively cooling gas.

These filaments can be traced to two distinct physical processes (\citealt{Pope}), the drift (a hydrodynamical effect related to the displacement of the gas by the bubbles; \citealt{Darwin1953}) and the wake (trapping of gas in eddies at the bottom of buouyantly rising bubbles; \citealt{Yang2003}). The drift is visible as the main body of the filaments in Fig. \ref{fig:fig8}, while the wake is visible as the 'petals' at the end of the filaments in the last three snapshots, most clearly in the X-ray maps.



The first snapshot in the X-rays shows the base of the cavities enveloped by cooling gas. This is qualitatively similar to filaments of cool gas enveloping the base of X-ray cavities, as observed with ALMA in the Phoenix cluster (\citealt{Russell2017}). The filamentary structures trailing bubbles in the second and third snapshot (and in our simulations in general, regardless of the resolution or mass of the system) are qualitatively in agreement with observations that find filaments trailing X-ray cavities or radio bubbles (e.g. \citealt{Russell2016}, \citealt{Vantyghem2018}). Observations with ALMA have found that filaments trailing bubbles may be ubiquitous wherever AGN bubbles are present (e.g. \citealt{Olivares2019}, \citealt{Russell2019}). Observations in other wavelengths have revealed many more examples of this correlation (e.g. \citealt{Wilman2009}, \citealt{Salome2011}, \citealt{Tremblay2015}, \citealt{Maccagni2021}). In the Virgo cluster, a pair of filaments are visible in X-rays, and they are aligned with a pair of radio lobes (\citealt{Feigelson1987}, \citealt{Bohringer1995}, \citealt{Gatuzz2021}). Other observations also find X-ray filaments trailing X-ray cavities (e.g. \citealt{Gendron-Marsolais2017}), but these generally require long exposure times in order to resolve the filaments.

Simulations have been able to reproduce the uplift that results in these filaments (e.g. \citealt{Churazov2001}, \citealt{Revaz2008}, \citealt{Li2014}, \citealt{Brighenti2015}, \citealt{Qiu2019}, \citealt{Zhang2022}), although it is not clear how common a feature they are. We have performed simulations of constant-power jets and jet-inflated bubbles in an idealised ICM (Huško et al. in prep.), where we found that the filaments are present after any bubble-inflation event. We also found that they are energetically significant, and that the process of uplift of ambient gas significantly reduces the central density of the ICM. This provides an alternative mechanism of feedback (alongside gas heating through shocks). It has even been suggested that jet feedback may represent a self-driven cycle: one jet episode results in the uplift of dense filaments that eventually cool and fall onto the central galaxy, triggering another jet episode (\citealt{McNamara2016}). We do find that these filaments eventually fall back onto the centre, but we leave a study of their role in the feedback cycle for a future paper.

\subsection{Structure of the cold gas}

Our simulations of the most massive halo ($M_\mr{200}=10^{15}$ $\mr{M}_\odot$) feature significant changes in the direction of the SMBH spin vector, as can be inferred from jets being launched in various directions. The evolution of the spin is primarily tied to the properties of the cold gas ($T<2\times10^4$ K) surrounding the SMBH. We find that the cold gas is morphologically varied. At times, it takes the form of a relatively long-lived, rotationally-supported disc (e.g. the first and last snapshots, visible mostly in the zoomed-in temperature and X-ray maps in Fig. \ref{fig:fig8}). At other times, it is relatively clumpy, and can even be located far from the SMBH (third snapshot). These variations could be attributed to: i) the depletion of gas due to direct launching into the jet by the jet-launching algorithm, ii) the entrainment of gas into the jet, iii) the cooling of gas at large distances due to shock compression of gas induced by the jets, iv) the cooling of filaments drawn out by jet-inflated bubbles and v) poor sampling due to finite numerical resolution.

Observationally, it is not clear how ubiquitous cold gaseous discs are in massive galaxy clusters. For some galaxy clusters there is clear evidence of molecular gas discs (\citealt{Hamer2014}), while for others there is evidence of most of the molecular gas residing in precipitating filaments (\citealt{Crawford2005}). An analysis of a sample of clusters by \cite{Russell2019} suggests that there is a spectrum, with most clusters having both filaments and circumnuclear discs, with neither dominating. Newer observations with ALMA (\citealt{Nagai2019}) find that many of these discs may be unresolved in lower-resolution observations. 

\subsection{Driving of shocks and sound waves}

As an anisotropic energy injection mechanism, jets are expected to deposit a significant fraction of their energy near the axis along which they are launched. In our most massive galaxy cluster simulation, this is not necessarily a problem since jets can reorient fairly quickly. In our lower-mass systems, such reorientation does not occur. 



From the maps of the shock/sound wave tracer in Fig. \ref{fig:fig8} we see that jet launching is accompanied by ellipsoidal or spherical shocks and sound waves that propagate throughout the halo. The shocks from multiple jet episodes interact with each other in a complex way, producing ripples with a radial direction. This likely results from interference of waves from different jet episodes (or from the two sides of a single jet episode). Plumes tracing the jet material are visible due to strong shocking of the jet gas. Sound waves in our simulations do not heat the ICM, since it is relatively homogeneous (i.e. it does not feature realistic substructures, such as gas clumps, sloshing fronts, relics of accreted clusters, etc.), but they might do so in realistic zoom-in cosmological simulations (\citealt{Bambic}). 


These plots show that jets that are directed along one axis can drive significant shocks in other directions. This is not surprising; many simulations have found that a significant fraction (usually of order $50\%$) of jet energy is imparted to the medium fairly isotropically while the jet is active, by driving a bow shock that deposits energy through thermalization at all angles (e.g. \citealt{Weinberger2017}, \citealt{Bourne2017}, \citealt{Husko2022a}). After the jet is turned off, even more (if not all) of the previously injected energy is imparted to the ambient medium. This is consistent with our galaxy group and low-mass cluster simulations, where jets are launched almost perfectly along the $z-$axis, yet they successfully quench gas cooling and star formation in the haloes.

\subsection{Impact of jet feedback on profiles of gas-related quantities}

Observations indicate that, in terms of X-ray properties, galaxy clusters come in roughly two types: cool-core (CC) and non-cool core (NCC; \citealt{McNamara2000}, \citealt{Lewis2002}). In their outer regions these clusters are very similar (\citealt{Voit2005}), but in their centres, CC clusters exhibit a dip in temperature that can be a few times lower than the peak (\citealt{Peterson2003}). This distinction is also visible in entropy (\citealt{Hudson2010}), density (\citealt{Peterson2006}) and pressure (\citealt{Arnaud2010}) profiles. CC clusters have significantly shorter central cooling times, typically less than the Hubble time (\citealt{Voit2015}). Previous simulations have shown that the distinction between CC and NCC clusters can be explained as a result of AGN feedback (e.g. \citealt{Dubois2011}, \citealt{Pike2014}, \citealt{Prasad2015}, \citealt{Barnes2017}). 

\begin{figure*}
\includegraphics[width=1.01\textwidth, trim = 0 10 0 0]{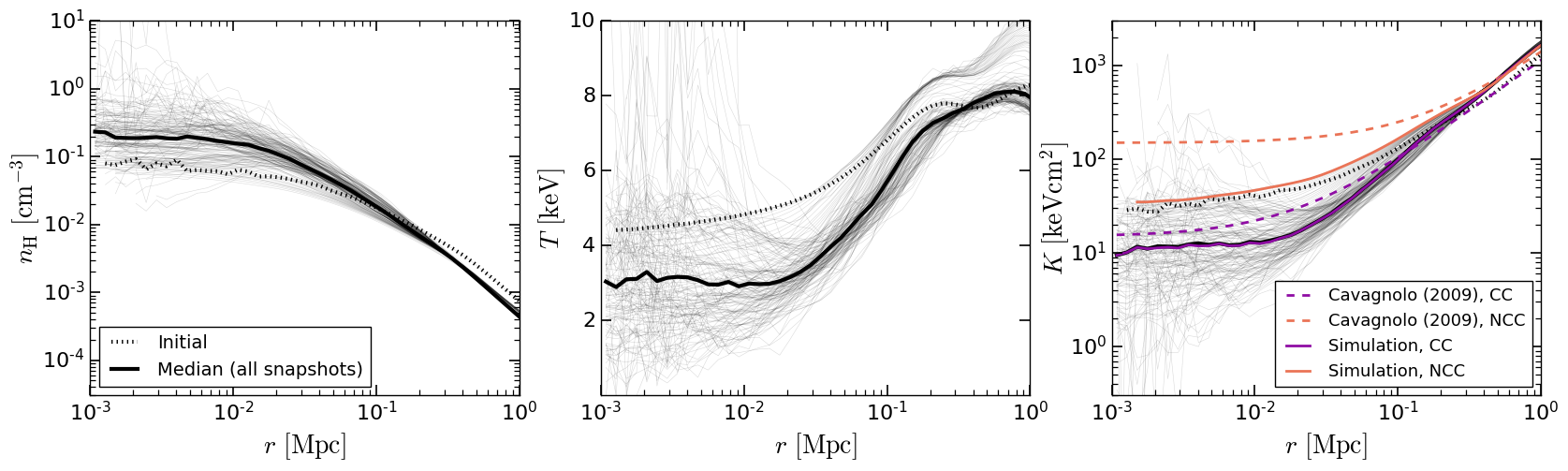}
\caption{Profiles of gas density, tempreature and entropy from our fiducal high-mass galaxy cluster simulation at medium resolution ($M_\mr{200}=10^{15}$ $\mr{M}_\odot$, $m_\mr{g}=6.4\times10^6$ $\mathrm{M}_\odot$, see Table \ref{tab:tab0} for details). The initial profiles are shown by dotted lines, while thick solid lines are the median profiles using individual snapshots, which are shown with thin solid lines. The purple and orange lines show median profiles for our simulated cluster when it is cool-core (CC) and non-cool-core (NCC), respectively. The cluster is classified as the former if its central entropy (within $r<10$ kpc) satisfies $K_0<30$ keVcm$^2$, and the latter if $K_0>30$ keVcm$^2$. This definition follows the observational sample of \protect\cite{Cavagnolo2009}; their median CC and NCC entropy profiles are shown with dashed purple and orange lines, respectively}.
\label{fig:fig11}
\end{figure*}

Fig. \ref{fig:fig11} shows the number density, temperature and entropy profiles of gas in our fiducial, medium-resolution simulation of the high-mass galaxy cluster ($M_\mr{200}=10^{15}$ $\mr{M}_\odot$). Outside 200 kpc, the median profiles over 16 Gyr of evolution are similar to the initial ones, indicating that feedback mostly has an effect on the region within that radius. There are some variations at different times at all radii, but these are related to the relaxation of the halo, which is most easily visible in the temperature profiles (shown here using a non-logarithmic axis), and also in an overall drop in the normalization of the density and entropy profiles.

Within $200$ kpc there are significant variations in the profiles between snapshots, due to active cooling and feedback in this region. The median temperatures within $r<10$ kpc reach values as low as a few times $10^6$ K ($T\approx0.1$ keV) and as high as $10^8$ K ($T\approx10$ keV), depending on whether the ICM is actively cooling or being heated by feedback. Entropies at the same time reach values as low as $K<1$ keVcm$^2$ and as high as $K=10^3$ keVcm$^2$. We also show median lines using all snapshots, as well as medians for when the cluster is considered CC or NCC for the entropy profiles, according to the definition of \cite{Cavagnolo2009} (a cluster is considered CC if its central entropy, $K_0$, measured within $r<10$ kpc, satisfies $K_0<30$ keVcm$^2$, otherwise it is NCC). We find that our simulated cluster is considered CC for almost its entire evolution, with the CC median lines and the overall medians being very similar. Most of the time the cluster is more CC than when initialized, with densities being higher and temperatures and entropies lower. Our CC median entropy profile agrees fairly well with the sample from \cite{Cavagnolo2009}, underestimating it by $\approx50\%$ in the centre. However, our NCC median entropy profile, comprised of only $\approx10$ snapshots, falls short of the observed NCC median from \cite{Cavagnolo2009} by a factor of $\approx4$. These differences may be in part due to the haloes in the sample of \cite{Cavagnolo2009} differing in mass from $M_{200}=10^{15}$ $\mathrm{M}_\odot$. It is also likely that more realistic, cosmological simulations, with sloshing due to mergers, would feature higher-entropy cores, both for CC and NCC clusters (\citealt{Ascasibar2006}, \citealt{ZuHone2010}).

\subsection{Impact of parameter variations on the cooling and feedback cycle of galaxies}

\begin{figure*}
\includegraphics[width=1.01\textwidth, trim = 0 10 0 0]{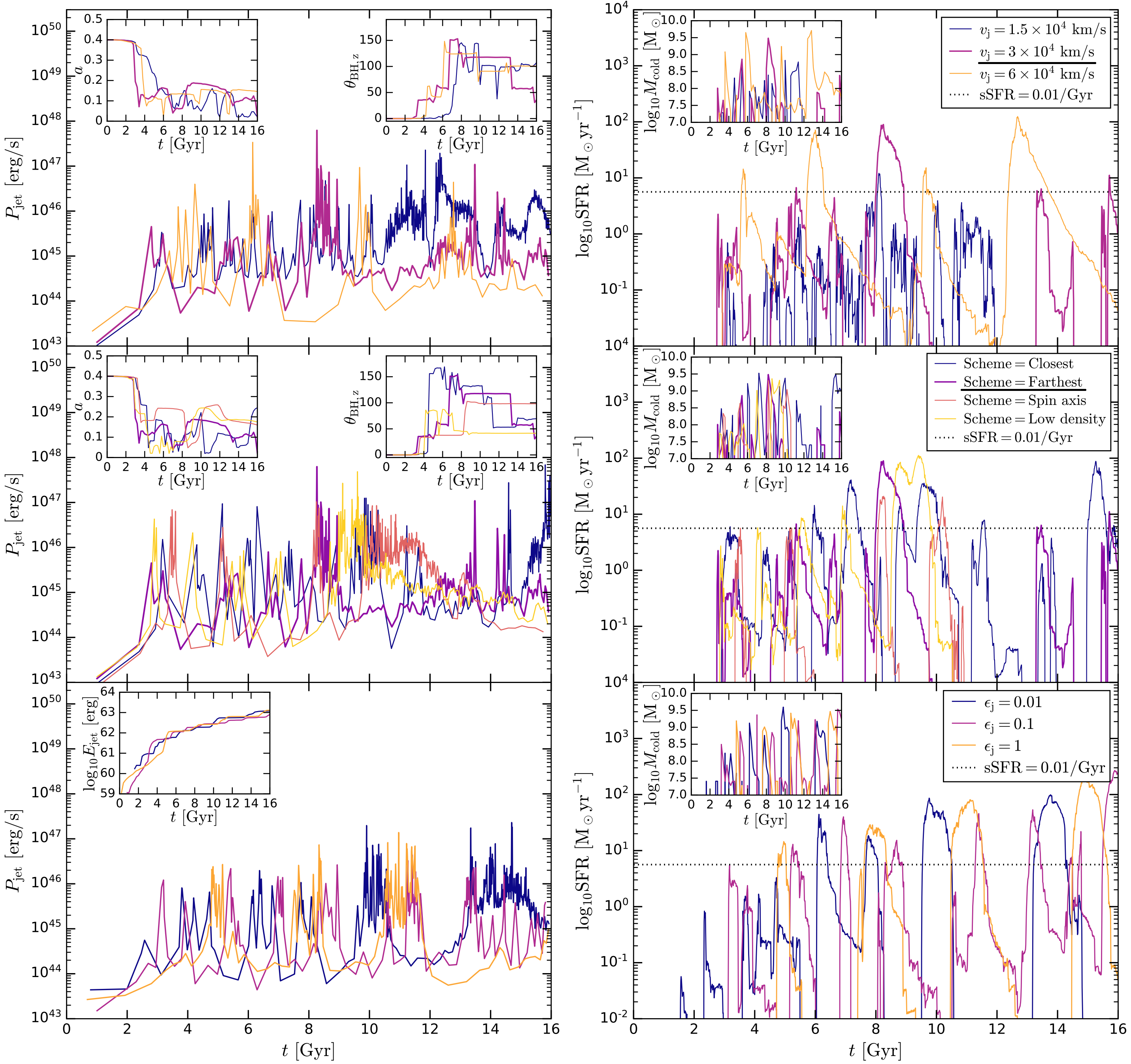}
\caption{Time dependence of the quenching/feedback process in the high-mass galaxy cluster simulations ($M_\mr{200}=10^{15}$ $\mr{M}_\odot$) with varying parameters/choices, as per the legends in the right-hand panels. In the top row we vary the jet launching velocity, and in the middle row the choice of which particles are kicked from the SMBH smoothing kernel. The bottom row shows a case with the jets fixed along the $z-$axis, with the jet power calculated using a fixed efficiency (see legend). The left-hand panels show the jet power, while the right-hand panels show the star formation rate. Insets in the left-hand panels show the magnitude of SMBH spin and the angle between the spin vector and $z-$axis in the top and middle panel, and the total injected jet energy in the bottom panel. The insets in the right-hand panels show the cold gas mass. The details of the fiducial case, relative to which these variations are made, are given in Table \ref{tab:tab0} (purple line in each panel and underlined parameter in each panel legend, with the exception of the bottom row). The dotted black lines represent the upper limit of the specific star formation rate required to classify a galaxy as quenched.}
\label{fig:fig12}
\end{figure*}

In \S~\ref{sec:sec4} we focused on general features of jets and the quenching process in all three systems that we simulated. We varied the initial SMBH spin, mass resolution and the central temperature in each case. In this section we present similar results, but for variations of other parameters or choices that we considered most significant; we discuss other variations in Appendix \ref{sec:app3} (where we find that they generally have little impact). These variations were all done for the high-mass galaxy cluster ($M_\mr{200}=10^{15}$ $\mr{M}_\odot$). In Fig. \ref{fig:fig12} we show the results of these variations for three different parameters/choices: the jet launching velocity, the scheme with which particles are kicked from the SMBH smoothing kernel, and finally a set of simulations where the jet direction is fixed along the $z-$axis, and the jet efficiency is also fixed in time. 


The top row of Fig. \ref{fig:fig12} shows results of varying the velocity with which particles are kicked from the SMBH smoothing kernel. In terms of jet power, higher launching velocities result in more episodic feedback, which is especially pronounced with the highest launching velocity we tested, $v_\mathrm{j}=6\times10^4\hspace{0.7mm}\mathrm{km}\hspace{0.2mm}\mathrm{s}^{-1}=0.2c$. This simulation has four cooling episodes (which may feature one or more distinct jet episode each) that last for $0.5-1.5$ Gyr. Between these cooling episodes, the jet power is very low. The high-velocity case is likely more episodic due to its more explosive nature (due to a larger launching velocity, stronger shocks occur as the jet is being decelerated, and at smaller distances). This difference results in lower jet powers in the minima between jet episodes; this is likely due to the presence of hotter gas in the centre of the halo. 

With lower launching velocities, the halo is heated more gently and at larger distances, since shocks occur at larger distances. This is a result of the jets being more mass and momentum loaded, since the total mass launched into a jet with a total energy $E_\mr{j}$ (which we consider constant for the purpose of this argument) is $M_\mathrm{j}=2E_\mathrm{j}/v_\mathrm{j}^2$, and the total momentum $p_\mathrm{j}=2E_\mathrm{j}/v_\mathrm{j}$. The jets are thus able to drill through the ICM more easily, if they are launched with lower velocities, until they have swept up approximately as much mass as the mass in the jets, which is roughly when they transition from the ballistic phase to the self-similar one (see e.g. \citealt{Kaiser2007} for a theoretical model, or \citealt{Husko2022a} for a confirmation of such behaviour in hydrodynamical tests). This transition roughly coincides with the scale where jets begin to experience strong shocking. Furthermore, since densities are smaller at larger radii, the shocks are also likely to be weaker.

The evolution of the SMBH spin magnitude and direction is similar in all three simulations, with perhaps the only exception being the somewhat more frequent changes in the spin for the lowest-velocity jet simulation, due to the jet being active throughout almost the entire simulation. Surprisingly, the peak star formation rates and cold gas masses are higher with larger launching velocities, which feature more explosive feedback. They also show more protracted decreases after their peaks during each episode. These differences are most likely due to explosive feedback being able to expel cold gas from the centre of the halo; the cold gas is then long-lived and star-forming until all of it is consumed (this behaviour was also found by \citealt{Nobels2022} with thermal AGN feedback). The lower-velocity cases feature stronger cold gas evacuation from the centre of the halo through the jet launching mechanism, leading to lower SFRs. By this we are not referring to entrainment, but rather that the jet launching algorithm chooses the cold gas to be launched into the jet. The reason this effect depends on the jet launching velocity is that the mass loading of the jet increases as the velocity decreases: $\dot{M}_\mr{j}=2P_\mr{j}/v_\mr{j}^2$.


When launching particles from the SMBH smoothing kernel, as part of our jet feedback implementation, a choice needs to be made as to which particles are launched (see \citealt{Chaikin2022}, in the context of stellar feedback). Our fiducial choice is to kick particles that are farthest from the SMBH on either hemisphere (relative to the spin vector). We compare this against kicking the closest two particles, as well as kicking the two particles closest to the spin axis (in terms of angular distance). We also compare against a case where the two particles with the lowest density are kicked, in an attempt to avoid launching cold gas into the jet. In Fig. \ref{fig:fig12} we show the effects of varying this choice. As we see, the consequences are minor but not negligible. 

The scheme with the two closest particles being kicked is overall most similar to our fiducial choice, where the two farthest particles are kicked. However, the jet powers, SFRs and cold gas masses are more variable and less episodic in that scheme, since the cold gas structure near the SMBH is more prone to being disrupted. The scheme where low-density gas is targeted is overall similar to the previous two, but results in quicker 'final' quenching by $t=10$ Gyr, due to a long-lived and strong cooling/jet episode, which is not easily disrupted since the launching scheme completely avoids the cold gas. The scheme where the gas closest to the spin axis is targeted appears to be most efficient at feedback. This scheme is most episodic and injects the energy required to quench cooling earlier than the other schemes. It completely quenches the halo by $t=10$ Gyr as well (at least out to the end of the simulation).

In the bottom row of Fig. \ref{fig:fig12} we show results from simulations with jets that are fixed along the $z-$axis, and that inject energy with a constant efficiency (that we vary). We find that there is surprisingly little difference among simulations with efficiencies varying by factors of 10. Lower efficiency cases have more variable cooling/jet episodes and shorter delays between the episodes. Overall, however, we find that the amounts of energy injected by the jets is similar in all three simulations. The same is true for the amount of star formation. This similarity is likely due to the self-regulated nature of the system (see e.g. \citealt{BoothSchaye2010}); the SMBHs may inject less energy in the beginning of a cooling flow if they have lower efficiencies, but this will quickly be compensated by more cooling (and higher SMBH accretion rates) until the jet heating rate becomes sufficient to offset the cooling.

\section{Summary and conclusions}
\label{sec:sec7}

In this paper we present a subgrid model for the spin evolution of SMBHs surrounded by unresolved thick accretion discs in hydrodynamical simulations of galaxy formation. This model is applicable for SMBHs accreting at low rates, as found in many galaxies in the local Universe, especially massive ones. Coupled with a spin-dependent jet efficiency formula based on recent general-relativistic magneto-hydrodynamical simulations, our model allows the study of self-consistent jet feedback in a realistic manner. Due to spin evolution, the model naturally results in jet reorientation, as well as changes in jet efficiency.

We have implemented our model into the SWIFT smoothed particle hydrodynamics code, and applied it to an idealised set-up that includes: 1) an external potential representing a dark matter halo 2) a central massive galaxy, 3) a realistic hot circumgalactic/intracluster medium in hydrostatic equilibrium and 4) a central black hole. We assume an accretion efficiency of $100\%$, so that the black hole accretion rate is equal to the Bondi accretion rate (no disc winds or other mass loss). The jet efficiencies that arise in our simulations are of order $1-10\%$, larger than most other similar simulations.

We have studied cases with three different dark matter halo masses: $10^{13}$ $\mr{M}_\odot$, $10^{14}$ $\mr{M}_\odot$ and $10^{15}$ $\mr{M}_\odot$. These set-ups represent typical systems where jet feedback is expected to be important, including galaxy groups and clusters. We have simulated these systems at different resolutions and with varying parameters. From these simulations we conclude the following:

\begin{itemize}
    \item Our jet feedback model is successful in quenching star formation in central galaxies across the mass scale, and with differing parameters related to jet feedback and initial conditions. Quenching is always achieved, but details of the feedback can depend on choices such as jet launching velocity and which particles are kicked from the SMBH smoothing kernel.
    \item The details of jet feedback are most sensitive to the mass of the system (as measured through the halo mass). In the $M_\mr{200}=10^{13}$ $\mr{M}_\odot$ case (a typical elliptical galaxy in a galaxy group), an initial strong cooling flow and jet episode leads to quenching within $0.5-2$ Gyr. This galaxy remains quenched for a further 6 Gyr. A weak, constant-power jet is fed directly from the halo of hot gas. In the largest system that we simulate, the high-mass galaxy cluster ($M_\mr{200}=10^{15}$ $\mr{M}_\odot$), the central galaxy experiences multiple cycles of cooling and jet activity. Jets fed by accretion from cold gas dominate in this system. In the intermediate-mass case ($M_\mr{200}=10^{14}$ $\mr{M}_\odot$), representing a low-mass galaxy cluster, cold gas accretion typically dominates, with hot halo accretion being sufficient to keep the halo quenched only if jet efficiencies are very high, of order $100\%$.
    \item At fixed halo mass, we find that the results are most sensitive to the initial central temperature of the gaseous halo. We find the strongest jet activity (with jet powers a few times $10^{47}$ $\mathrm{erg}\hspace{0.3mm}\mathrm{s}^{-1}$) in our high-mass galaxy cluster ($M_\mr{200}=10^{15}$ $\mr{M}_\odot$), if it is initialized as a strong cool-core cluster. Cold gas masses in this case reach peak values of $10^{10}$ $\mr{M}_\odot$ and star formation rates reach peak values of a few times $100$ $\mr{M}_\odot\mr{yr}^{-1}$, in agreement with observations. Periods of such high cold gas masses, star formation rates and jet powers can last anywhere from $0.1$ Gyr to 1 Gyr, depending on the jet efficiencies (i.e. spin) and the details of the system in question.
    \item In the galaxy group ($M_\mr{200}=10^{13}$ $\mr{M}_\odot$) and low-mass cluster ($M_\mr{200}=10^{14}$ $\mr{M}_\odot$), the cooling flows that develop do not lead to cold gas mass reservoirs large enough, or long-lived enough, to lead to significant SMBH accretion or spin evolution. Significant evolution of SMBH spin (both in terms of magnitude and direction) occurs in the $M_\mr{200}=10^{15}$ $\mr{M}_\odot$ system. The accretion is chaotic and not well-aligned with the $z-$axis, with the gas sometimes forming a circumnuclear disc, and at other times clumps that may appear at large distances from the centre of the halo ($>10$ kpc).
    \item Compared to simulations using thermal AGN feedback in the same set-up, performed by \cite{Nobels2022}, we find that jets are more efficient at quenching the galaxies. They lead to overall less star formation and cold gas, as well as more rapid shutoff in star formation during a given cooling flow. Compared to thermal feedback, the jets are able to quench haloes with lower initial central gas temperatures. The cooling and feedback cycle is periodic in the case with thermal feedback, unlike the jet feedback case, where the time elapsed between cooling episodes is less predictable. This is likely due to jet efficiencies that vary during a given simulation.
    \item The inflation of jet lobes/bubbles is always followed by the uplift of low-entropy gas from the centre of the gaseous halo. This gas forms dense, cooling filaments, in agreement with observations that suggest that these filaments are ubiquitous in galaxy clusters with evidence of jet activity.
\end{itemize}

Our simulations of the group and low-mass cluster regimes ($M_\mr{200}=10^{13}$ $\mr{M}_\odot$ and $M_\mr{200}=10^{14}$ $\mr{M}_\odot$) featured almost no spin evolution, which means that the jet efficiency and direction were effectively fixed. In addition, we performed some simulations of the high-mass galaxy cluster ($M_\mr{200}=10^{15}$ $\mr{M}_\odot$) with the jet efficiency fixed at a few different values and the jet direction fixed along the $z-$axis. In all these simulations, successful quenching was achieved. These results indicate that variations of the jet efficiency and direction due to BH spin evolution may not be important if the main goal is to quench galaxies. However, some secondary effects are probably lost (e.g. non-periodicity of cooling flows). In a follow-up paper, we plan to investigate the importance of varying jet efficiencies and directions in detail, using the same set-up as in this paper. In the same paper we will compare jet feedback with the thermal feedback mode used in the EAGLE simulations.

In the future we also plan to extend our analysis to idealised set-ups representing different physical systems where jets may be important, e.g. disc galaxies or galaxy mergers. We will also perform cosmological zoom-in simulations in order to study jets in a more realistic, cosmological context. Eventually, we plan to perform large-volume cosmological simulations with jets as a feedback mechanism.

\section*{Acknowledgements}
We thank the referee for a helpful report. The research in this paper made use of the SWIFT open-source simulation code (\url{http://www.swiftsim.com}, \citealt{Schaller2018})
version 0.9.0. The swiftsimio Python library was used to analyze and visualize the data from the simulations (\citealt{Borrow2020_swiftsimio}, \citealt{Borrow_2021_swiftsimio}). The work has been performed under the Project HPC-EUROPA3 (INFRAIA-2016-1-730897), with the support of the EC Research Innovation Action under the H2020 Programme; in particular, F.H. gratefully acknowledges the support of the Leiden Observatory and the computer resources and technical support provided by SURFsara, the Dutch national HPC facility. F. H. would like to acknowledge support from the Science Technology Facilities Council through a CDT studentship (ST/P006744/1), and the STFC consolidated grant ST/T000244/1. This work used the DiRAC@Durham facility managed by the Institute for Computational Cosmology on behalf of the STFC DiRAC HPC Facility (www.dirac.ac.uk). The equipment was funded by BEIS capital funding via STFC capital grants ST/K00042X/1, ST/P002293/1, ST/R002371/1 and ST/S002502/1, Durham University and STFC operations grant ST/R000832/1. DiRAC is part of the National e-Infrastructure.

\section*{Data availability}

The data underlying this article will be provided upon request to the corresponding author. The code and initial conditions used to generate the data are available online (\href{https://github.com/SWIFTSIM/swiftsim}{https://github.com/SWIFTSIM/swiftsim}).

\bibliographystyle{mnras}
\bibliography{mnras_template} 

\clearpage
\newpage
\mbox{}

\setcounter{page}{1}

\appendix

\section{Specific angular momentum at the innermost stable circular orbit}
\label{sec:app1}

The radius of the innermost stable circular orbit (ISCO) is given by $R_\mr{ISCO}=r_\mathrm{ISCO} R_\mr{G}$, where $R_\mr{G}=M_\mr{BH}G/c^2$ and $r_\mathrm{ISCO}$ is the dimensionless radius of the ISCO. $r_\mathrm{ISCO}$ depends on spin as
\begin{equation}
    r_\mathrm{ISCO}=3+Z_2\mp\sqrt{(3-Z_1)(3+Z_1+2Z_2)},
\label{eq:Lambda}
\end{equation}
where the minus and plus sign are for prograde and retrograde accretion, respectively, and $Z_1$ and $Z_2$ are functions of spin given by
\begin{equation}
    Z_1(a)=1+\Big(1-a^2\Big)^{1/3}\Big[\big(1+\vert a \vert\big)^{1/3}+\big(1-\vert a\vert\big)^{1/3}\Big]
\label{eq:Z1}
\end{equation}
and
\begin{equation}
    Z_2(a)=\sqrt{3a^2+Z_1}.
\label{eq:Z2}
\end{equation}
The specific angular momentum at the ISCO is given by $L_\mr{ISCO}=M_\mr{BH}G\ell_\mr{ISCO}/c$, where $\ell_\mr{ISCO}$ is a dimensionless function of spin:
\begin{equation}
    \ell_\mr{ISCO}(a)=\frac{2}{3\sqrt{3}}\Big(1+2\sqrt{3r_\mathrm{ISCO}-2}\Big).
\label{eq:l_ISCO}
\end{equation}

\section{Black hole spin alignment timescale}
\label{sec:app2}

Our procedure for modeling SMBH spin builds on the approach of several previous studies (\citealt{Volonteri2007}, \citealt{King2008}, \citealt{Fanidakis2011}, \citealt{Griffin2019a}), with the main difference being that those studies focused on thin, radiatively efficient discs. This method is similar in many aspects to that implemented by \cite{Fiacconi2018} and \cite{Talbot2020}, which also modeled the thin disc. However, we do not explicitly model the differential equation for the evolution of the SMBH spin direction due to LT torques between the accretion disk and the SMBH (e.g. \citealt{Martin2007}). 

Furthermore, \cite{Fiacconi2018} measure the inflow of mass and angular momentum onto their sink particle, representing the SMBH and accretion disc system (without assuming e.g. Bondi accretion). This allows them to model the size, as well as total mass and angular momentum of their subgrid accretion disc (assuming some surface density profile, as well as Keplerian rotation). In turn this leads to the mass and angular momentum accretion rates onto the SMBH, using a system of equations that couple the SMBH with the large-scale accretion disc. These accretion rates onto the SMBH are in general different from the inflow rates onto the sink particle.

We instead assume Bondi accretion, and that the accretion rate onto the SMBH is equal to the Bondi rate at all times. For the angular momentum evolution, we use only the \textit{direction} of angular momentum measured in the smoothing kernel of the SMBH. While it is beyond the scope of this paper to compare our approach with the full approach of \cite{Fiacconi2018}, we will compare the alignment timescale in our approach to that where LT torques are explicitly included in the angular momentum evolution equation. We do this for the radiatively-efficient thin disc assuming that gas pressure dominates over radiation pressure, and that free-free absorption dominates in the opacity (region C from \citealt{ShakuraSunyaev1973}). We make this choice since \cite{Fiacconi2018} studied this accretion regime, and since they use the derivation of \cite{Martin2007} to model the terms from the LT torque that contribute to the angular momentum evolution equation, which is applicable only for Keplerian rotation and thin discs. Note that while we assume the thick accretion disc in this paper, a comparison between the differential equation approach and the incremental accretion approach in any accretion regime should serve as a general validation of the incremental accretion approach (with the caveat that the differential equation itself should change depending on accretion regime).

In the \cite{Fiacconi2018} approach, the evolution of angular momentum is given by the equation
\begin{equation}
\begin{aligned}
    \frac{\mathrm{d}\mathbf{J}_\mathrm{BH}}{\mathrm{d}t} = &  L_\mathrm{ISCO}\dot{M}_\mathrm{BH,0}\mathbf{\hat{J}_\mathrm{BH}}- \\ 
    &\frac{J_\mathrm{BH}}{\tau_\mathrm{GM}}\big[ \tilde{K}_1 \mathbf{\hat{J}_\mathrm{BH}}\times \mathbf{\hat{J}_\mathrm{d}} +\tilde{K}_2\mathbf{\hat{J}_\mathrm{BH}}\times(\mathbf{\hat{J}_\mathrm{BH}}\times\mathbf{\hat{J}_\mathrm{d}})  \big],
\label{eq:Fiacconi}
\end{aligned}
\end{equation}
where $\mathbf{J}_\mathrm{BH}$ is the angular momentum vector of the BH, $\mathbf{\hat{J}_\mathrm{BH}}$ is its direction, $\mathbf{\hat{J}_\mathrm{d}}$ is the direction of the large-scale accretion disc surrounding the black hole (outside the warp radius), $\tilde{K}_1=\sin(\pi/7)$, $\tilde{K}_2=\cos(\pi/7)$ and $\tau_\mathrm{GM}$ is a gravito-magnetic timescale which is given by
\begin{equation}
    \tau_\mathrm{GM}=0.17\hspace{0.3mm}\mathrm{Myr}\hspace{0.3mm}\bigg(\frac{M_\mathrm{BH}}{10^6\hspace{0.3mm}\mathrm{M}_\odot}\bigg)^{-2/35}\dot{m}^{-32/35}\vert a \vert^{5/7}.
\label{eq:tau_gm}
\end{equation}
for the thin accretion disc. The first term in equation (\ref{eq:Fiacconi}) corresponds to accretion onto the SMBH, and is identical to what we assume. Other terms represent the effects of LT torques in a warped accretion disc. Their form, including the forms of $\tilde{K}_1$, $\tilde{K}_2$ and $\tau_\mathrm{GM}$, follow from the derivation in \cite{Martin2007} under the assumption that the surface density scales as $\Sigma(R)\sim R^{-3/4}$, as appropriate for region C of the \cite{ShakuraSunyaev1973} thin disc, and that the tilt angle between the SMBH spin vector and the outer accretion disc is small. 

The first term in the parentheses in equation (\ref{eq:Fiacconi}) causes precession, whereas the second term leads to alignment. If we define the alignment timescale such that the alignment term in equation (\ref{eq:Fiacconi}) takes the form $(\mathrm{d}\mathbf{J}_\mathrm{BH}/\mathrm{d}t)_\mathrm{align}=-(J_\mathrm{BH}/\tau_\mathrm{align})\mathbf{\hat{J}_\mathrm{BH}}\times(\mathbf{\hat{J}_\mathrm{BH}}\times\mathbf{\hat{J}_\mathrm{d}})$, the alignment timescale in this differential equation approach can be written as 
\begin{equation}
\begin{aligned}
    \tau_\mathrm{align,diff.\hspace{0.15mm}eqn.}&=\frac{\tau_\mathrm{GM}}{\cos(\pi/7)}=\\
    &=0.19\hspace{0.3mm}\mathrm{Myr}\hspace{0.3mm}\bigg(\frac{M_\mathrm{BH}}{10^6\hspace{0.3mm}\mathrm{M}_\odot}\bigg)^{-2/35}\dot{m}^{-32/35}\vert a \vert^{5/7}.
\label{eq:tau_align_diff_eqn}
\end{aligned}
\end{equation}

In the approach we use in this paper (we refer to this as the warp increment approach), we use only the first term of equation (\ref{eq:Fiacconi}) to evolve the magnitude of angular momentum (i.e. spin). To evolve its direction, at the end of every time-step we assume that the new direction of angular momentum matches that of $\mathbf{J}_\mathrm{BH}+\Delta\mathbf{J}_\mathrm{warp,tot}$, where $\Delta\mathbf{J}_\mathrm{warp,tot}=\Delta J_\mathrm{warp,tot}\mathbf{\hat{J}}_\mathrm{d}$ is the total angular momentum of all of the warp increments consumed over the time-step. This material is originally aligned with the outer accretion disc (in the direction of $\mathbf{\hat{J}}_\mathrm{d}$), but we assume that it is (counter-)aligned with the SMBH angular momentum vector through LT torques. In the process, the SMBH is also torqued in the opposite direction. The magnitude of the consumed angular momentum is given by $\Delta J_\mathrm{warp,tot}=N_\mathrm{warp}J_\mathrm{warp}$, where $N_\mathrm{warp}=\Delta M_0/M_\mathrm{warp}$ is the number of warp increments consumed over the time-step, given the mass to be consumed, $\Delta M_0$, and $J_\mathrm{warp}$ is the angular momentum of one warp increment. Equally, one can view this as the SMBH being torqued by gas with a specific angular momentum of $L_\mathrm{warp}=J_\mathrm{warp}/M_\mathrm{warp}$ as it accretes. For the general case of $\Sigma(R)\sim R^{p}$, we have $L_\mathrm{warp}=(2+p)/(5/2+p)\times\sqrt{M_\mathrm{BH}GR_\mathrm{warp}}$, where $R_\mathrm{warp}$ is the warp radius (see appendix in \citealt{Fiacconi2018} for expression for region C of the thin disc). For the case of interest here, $\Sigma \sim R^{-3/4}$, and thus the numerical factor evaluates to $5/7$. Using the above relations, we can define a similar alignment timescale to the one in the differential equation approach, namely
\begin{equation}
\begin{aligned}
    \tau_\mathrm{align,warp}&=\bigg(\frac{J_\mathrm{BH}\Delta t }{\Delta J_\mathrm{warp,tot}}\bigg)=\\
    &=0.21\hspace{0.3mm}\mathrm{Myr}\hspace{0.3mm}\bigg(\frac{M_\mathrm{BH}}{10^6\hspace{0.3mm}\mathrm{M}_\odot}\bigg)^{-2/35}\dot{m}^{-32/35}\vert a \vert^{5/7}.
\label{eq:tau_align_warp}
\end{aligned}
\end{equation}
This alignment timescale differs from the differential equation one by an order-unity numerical factor. However, note that they are only directly comparable in the case that the SMBH angular momentum vector, and the angular momentum vector of the outer accretion disc, are perpendicular. Furthermore, we generally do not expect the two timescales to be the same, since these two approaches appear inherently different.

In order to compare the two approaches in more detail, we use a very simple test set-up. We assume a SMBH mass of $M_\mathrm{BH}=10^7$ $\mathrm{M}_\odot$, spin of $a=0.5$ directed along the $z-$axis, and constant accretion rate of $\dot{m}=0.1$. We assume that the angular momentum of the accretion disc on large scales, $\mathbf{\hat{J}}_\mathrm{d}$, is directed along the $x-$axis. Our aim is to compare how long it takes for the spin vector to be redirected in the direction of $\mathbf{\hat{J}}_\mathrm{d}$ in the two approaches. With our assumed parameters, we find that both the mass and spin magnitude remain very close to their initial values over the timescale of realignment, which is in this case of order a few Myr. We evolve the system using $100$ time steps between $t=0$ Myr and $t=10$ Myr in both approaches.

\begin{figure}
\includegraphics[width=1.01\columnwidth, trim = 0 10 0 0]{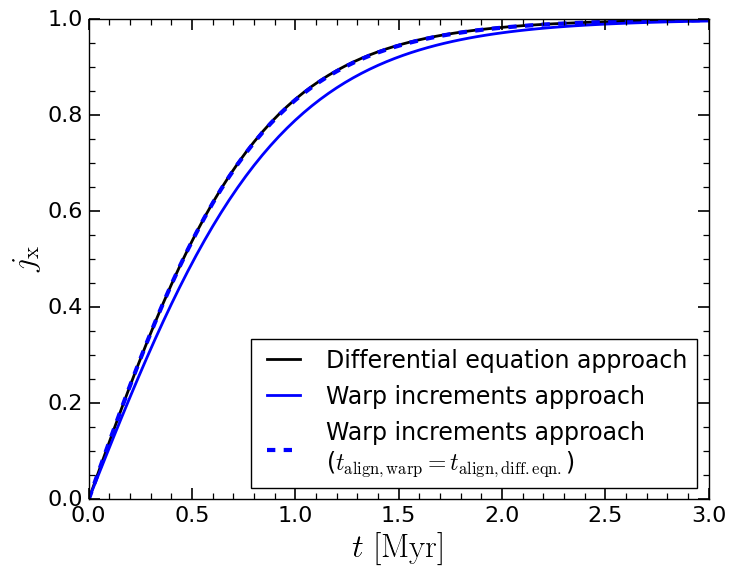}
\caption{The alignment of an initially misaligned SMBH spin vector, calculated using two different approaches. We show the $x-$ component of the angular momentum direction vector, i.e. the component in the direction of the outer accretion disc. The initial SMBH spin is $a=0.5$, and it is directed along the $z-$axis. The SMBH mass is $M_\mathrm{BH}=10^7$ $\mathrm{M}_\odot$, and the SMBH is accreting with a constant $\dot{m}=0.1$. Black lines show the predicted alignment from the differential equation approach, using equations (\ref{eq:Fiacconi}) and (\ref{eq:tau_align_diff_eqn}), while blue lines show predictions using our warp increment approach, with equation (\ref{eq:tau_align_warp}). The solid blue line shows the predicted alignment using the derived warp alignment timescale (equation \ref{eq:tau_align_warp}), while the dashed blue line shows the prediction from the same approach, but using the alignment timescale from the differential equation approach, given by equation (\ref{eq:tau_align_diff_eqn}). The agreement between the black and dashed blue lines shows that the two approaches are (almost) equivalent.}
\label{fig:figA1}
\end{figure}

In Fig. \ref{fig:figA1} we show the $x-$component of the angular momentum direction of the SMBH, i.e. the component in the direction of the outer accretion disc. The alignment is complete within $2-3$ Myr in both approaches, but it appears to be somewhat slower in our approach, using warp increments. We have attempted our approach with the warp angular moomentum boosted slightly so that the alignment time-scales is exactly equal to the one in the differential equation approach (i.e. so that the numerical factor in equation \ref{eq:tau_align_warp} evaluates to 0.19 instead of 0.21, to match equation \ref{eq:tau_align_diff_eqn}). This change leads to the two approaches of modeling SMBH spin alignment agreeing perfectly, showing that they are compatible.

Since we assume a thick disc in the model presented in this paper, rather than the thin disc, it is worth addressing the alignment timescale for that accretion regime. The surface density in the self-similar thick solution of \cite{Narayan1995} scales as $\Sigma(R) \sim R^{-1/2}$, so $L_\mathrm{warp}=(3/4)\Omega_0\sqrt{M_\mathrm{BH}GR_\mathrm{warp}}$, where we have also assumed that the orbital velocities are a fraction $\Omega_0$ of their Keplerian values (see discussion in \S \ref{sec:prograde_retrograde}). Defining the alignment timescale in the same way as in equation (\ref{eq:tau_align_warp}) leads to
\begin{equation}
    \tau_\mathrm{align,warp,thick} = \frac{4M_\mr{BH}}{3\dot{M}_\mr{BH,0}}\vert a\vert\Omega_0\bigg(\frac{384\vert a \vert}{25 (H/R)^2}\bigg)^{-1/5},
\label{eq:tau_align_warp_thick}
\end{equation}
where we have used equation (\ref{eq:r_warp_adaf}) for the warp radius of the thick disc. Using our fiducial values for the parameters leads to $\tau_\mathrm{align,warp,thick}= 0.63\vert a \vert^{4/5}(M_\mathrm{BH}/\dot{M}_\mathrm{BH,0})$. This shows that the alignment timescale of a thick disc is of order the growth timescale of the SMBH. This is generally much slower than in the thin disc case, owing to the vast difference in the warp radii (several vs. thousands of $R_\mathrm{G}$, respectively), and therefore even larger differences in warp angular momenta. For a value of $a=0.25$ (the equilibrium spin value, relevant in our high-mass galaxy cluster simulations), the alignment timescale is significantly shorter than the growth timescale, $\tau_\mathrm{align,warp,thick}\approx 0.2(M_\mathrm{BH}/\dot{M}_\mathrm{BH,0})$, indicating that the SMBH can be redirected without having to significantly grow its mass.

\section{Additional parameter variations}
\label{sec:app3}

Here we provide results on and discuss variations of different parameters related to jet feedback and our setup. These simulations were all of our high-mass galaxy cluster ($M_\mathrm{200}=10^{15}$ $\mathrm{M}_\odot$), simulated at medium resolution ($m_\mathrm{g}=6.4\times10^6$ $\mathrm{M}_\odot$).

\subsection{Direction of jet launching: opening angle and inclination}

\begin{figure*}
\includegraphics[width=1.01\textwidth, trim = 0 10 0 0]{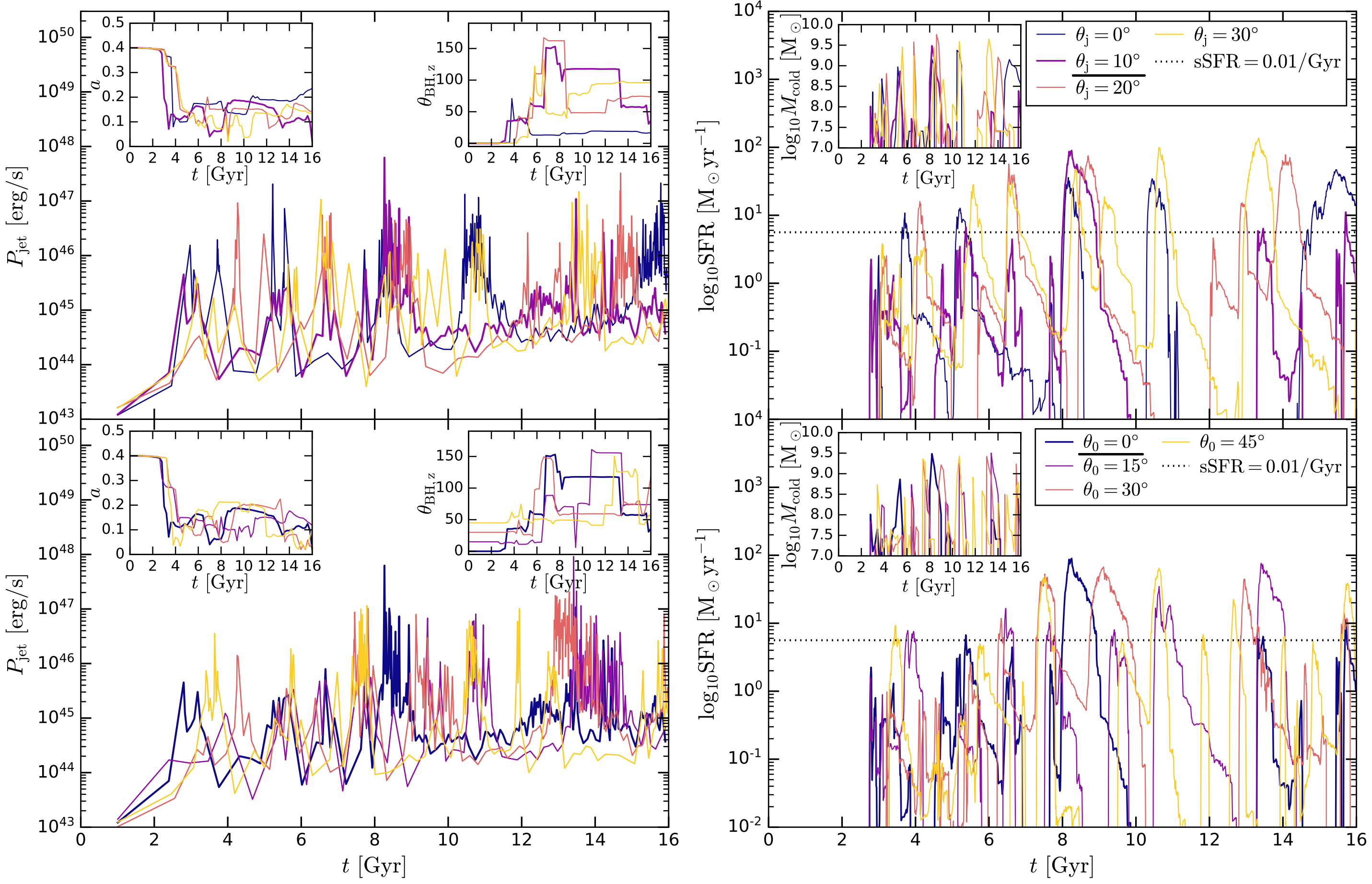}
\caption{Time dependence of the quenching/feedback process in the high-mass galaxy cluster simulations ($M_\mathrm{200}=10^{15}$ $\mathrm{M}_\odot$) with varying half-opening angles of the jet (top) and initial inclinations of the SMBH spin vector, relative to the angular momentum of the halo (bottom). The left-hand panels show the jet power, while the right-hand panels show the star formation rate. Insets in the left-hand panels show the magnitude of SMBH spin and the angle between the spin vector and $z-$axis. The insets in the right-hand panels show the cold gas mass. The fiducial simulation is the one with half-opening angle $\theta_\mathrm{j}=10\degree$ and inclination $\theta_0=0\degree$ (purple and blue line, respectively, in the top and bottom panels); other details of the fiducial setup are given in Table \ref{tab:tab0}. The dotted black lines represent the upper limit of the specific star formation rate required to classify a galaxy as quenched.}
\label{fig:figA2}
\end{figure*}

In the top row of Fig. \ref{fig:figA2} we show the effects of varying the half-opening angle of the jets (the fiducial value being $\theta_\mathrm{j}=10\degree)$. We find that all relevant quantities are very similar, even with a ballistic jet ($\theta_\mr{j}=0\degree$). This is somewhat surprising, given the fact that jets with smaller opening angles are able to reach larger distances (e.g. \citealt{Kaiser2007}, \citealt{Husko2022a}). The only visible difference we find is that the ballistic jet appears to feature less reorientation relative to the initial direction (visible in the plot of the misalignment angle). In addition, the spin of the SMBH takes on larger values, on average. In the bottom row of Fig. \ref{fig:figA2} we show results of varying the angle between the angular momentum vector of the halo and the initial SMBH spin vector of the SMBH (the fiducial angle being $\theta_0=0\degree$). The results are again very similar in terms of jet power, SFR and cold gas. There are no discernible differences among these quantities with different misalignment angles. This is likely indicative of the chaotic nature of these simulations.

\subsection{Impact of ICM properties}

\begin{figure*}
\includegraphics[width=1.01\textwidth, trim = 0 10 0 0]{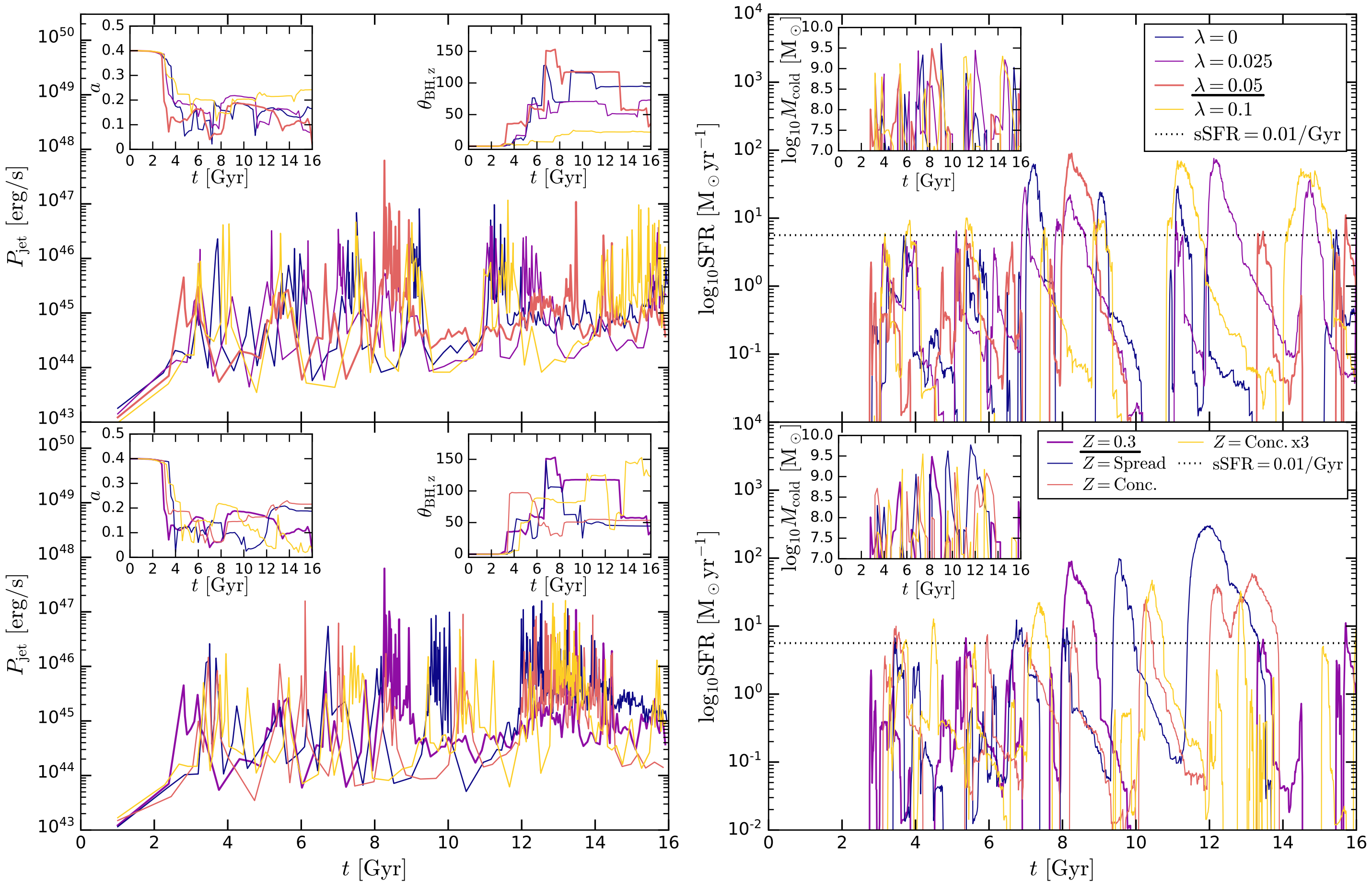}
\caption{Time dependence of the quenching/feedback process in the high-mass galaxy cluster simulations ($M_\mathrm{200}=10^{15}$ $\mathrm{M}_\odot$) with varying spin parameters (top) and metallicity profiles of the ICM (bottom). The left-hand panels show the jet power, while the right-hand panels show the star formation rate. Insets in the left-hand panels show the magnitude of SMBH spin and the angle between the spin vector and $z$-axis. The insets in the right-hand panels show the cold gas mass. The fiducial simulation is the one with an ICM spin parameter of $\lambda_\mathrm{g}=0.05$ and constant metallicity $Z=0.3Z_\odot$ (orange and purple lines, respectively, in the top and bottom panels); other details of the fiducial setup are given in Table \ref{tab:tab0}. The dotted black lines represent the upper limit of the specific star formation rate required to classify a galaxy as quenched.}
\label{fig:figA3}
\end{figure*}

In the top row of Fig. \ref{fig:figA3} we show results of varying the spin parameter of the gaseous halo of our galaxy cluster. Our fiducial value is $\lambda_\mathrm{g}=0.05$, and here we show cases with values that are equal to half and double our fiducial, as well as no net rotation of the halo. The differences between these cases are minor, especially in terms of jet power, cold gas or SFR. The case with largest rotation ($\lambda_\mathrm{g}=0.1$) features on average somewhat higher spins and is never significantly misaligned from the $z$ axis. Overall, these results imply that the rotation of the halo does not play a very significant role, and that the behaviour of the cooling gas is chaotic. These conclusions should be treated carefully, however, since these simulations were performed with a resolution of $6.4\times 10^6$ $\mr{M}_\odot$, which means that the cold gas is sampled with $100-1000$ particles at most.

In the bottom row of Fig. \ref{fig:figA3} we show variations in metallicity profiles of gas. We compare our fiducial case, with a constant metallicity of $0.3Z_\odot$, against a case with a 'spread' metallicity profile in accord with observations, $Z(r)=Z_\odot/(1+r/20\hspace{0.3mm}\mr{kpc})^{0.26}$ (which falls from solar values to $0.3Z_\odot$ at the virial radius), a 'concentrated' profile $Z(r)=Z_\odot/(1+r/20\hspace{0.3mm}\mr{kpc})^{0.5}$ (which falls to $0.3Z_\odot$ by 100 kpc), and a similar one ('3xconcentrated') that has a central metallicity of $3Z_\odot$, but also falls to $0.3Z_\odot$ by $100$ kpc. The 'spread' case features the highest SFR of all the simulations. The two other cases do not show significant differences relative to the fiducial case, despite the central metallicity being $\approx3$ and $10$ times higher. This is likely because the higher metallicities cause faster cooling only for $T<$ a few $\times$ $10^6$ K. However, gas that has already cooled down to such temperatures is already rapidly cooling, so the higher metallicities have an effect only on how quickly this portion of cooling occurs. The cooling gas spends the vast majority of its time at $T>10^7$ K, so metallicity plays only a small role in these simulations.

\subsection{Impact of other variations}

\begin{figure*}
\includegraphics[width=1.01\textwidth, trim = 0 10 0 0]{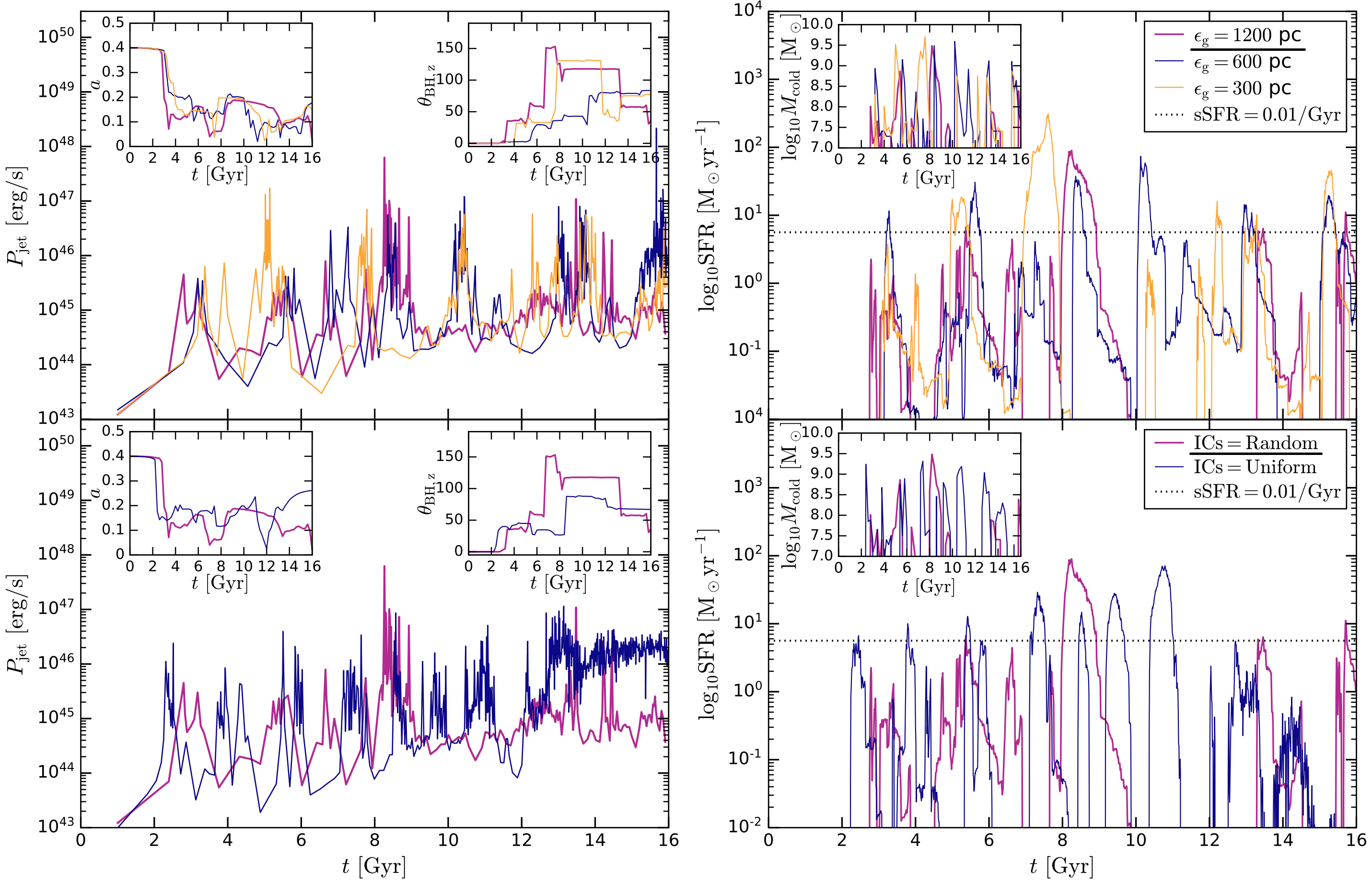}
\caption{Time dependence of the quenching/feedback process in high-mass galaxy cluster simulations ($M_\mathrm{200}=10^{15}$ $\mathrm{M}_\odot$) with varying gravitational softening lengths at fixed mass resolution (top) and particle position seeding (bottom). The left-hand panels show the jet power, while the right-hand panels show the star formation rate. Insets in the left-hand panels show the magnitude of SMBH spin and the angle between the spin vector and $z$-axis. The insets in the right-hand panels show the cold gas mass. The fiducial simulation is the one with a gravitational softening length of of $\epsilon_\mathrm{g}=1.2$ kpc and random particle seeding (purple line in each panel); other details of the fiducial setup are given in Table \ref{tab:tab0}. The dotted black lines represent the upper limit of the specific star formation rate required to classify a galaxy as quenched.}
\label{fig:figA4}
\end{figure*}

In the top row of Fig. \ref{fig:figA4}, we show the effects of reducing the gravitational softening length from $1200$ pc, our fiducial choice, to $600$ and $300$ pc. These simulations are very similar. The one with the lowest softening length reaches the highest peak in SFR, but that is possibly incidental and due to the chaotic nature of these massive galaxy cluster simulations. The spin evolution in the three simulations is remarkably similar, despite the fact that the evolution of the misalignment between the spin vector and $z-$axis is not. Reducing the softening length does not appear to result in more coherent rotating, cold gas discs near the SMBH.

In the bottom row of Fig. \ref{fig:figA4} we compare two simulations with different initial conditions: random seeding (our fiducial choice) versus a uniform cubic lattice of particles, that is then stretched to produce the desired density profile. The differences are clearly visible. With uniform initial conditions, the gas cools more rapidly and there are generally more frequent jet/star formation episodes. This is likely because cooling caused by thermal instabilities from random seeding is not present, so the eventual cooling flows are stronger and harder to suppress (we find that overall more jet energy is injected in this simulation). Despite the lack of initial randomness, even the uniform simulation results in spindown and misalignment of the spin vector (with turbulence possibly being caused by jet feedback itself).

\bsp	
\label{lastpage}
\end{document}